\documentclass{aa}
\usepackage{times}
\usepackage{array}
\usepackage{epsfig}
\usepackage{amssymb}
\usepackage{amsmath}
\usepackage{longtable}
 \def\mso{\,\mathrm{M}_\odot}                                                  
 \def\rso{\,{\rm R}_\odot}                                                      
 \def\lso{\,{\rm L}_\odot}                                                      

 \def\h1{\hangindent=1.0truecm \hangafter=0}                                    
 \def\simle{\mathrel{\hbox{\rlap{\hbox{\lower4pt\hbox{$\sim$}}}\hbox{$<$}}}}    
 \def\simgr{\mathrel{\hbox{\rlap{\hbox{\lower4pt\hbox{$\sim$}}}\hbox{$>$}}}}    
 \def\msoy{\, \mso~{\rm yr}^{-1}}

 \def\hb{\hfill\break}
 \def\h2{\hb\noindent \hangindent=0.5cm \hangafter=1}

 \def\utw{\smash{\rlap{\lower5pt\hbox{$\sim$}}}}
 \def\udtw{\smash{\rlap{\lower6pt\hbox{$\approx$}}}}

\begin{document}

\newcommand{\Lsun}{{\mathrm{L}_{\odot}}}
\newcommand{\Msun}{{\mathrm{M}_{\odot}}}
\newcommand{\Rsun}{{\mathrm{R}_{\odot}}}

\authorrunning{S. Wellstein et al.}

\title{Formation of contact in massive close binaries}

\author{S. Wellstein\inst{1}, N. Langer\inst{1,2} \and H. Braun\inst{3}}

\institute{
Institut f\"ur Physik, Universit\"at
Potsdam, D--14415~Potsdam, Germany
\and
Astronomical Institute, Utrecht University, Princetonplein 5,
NL-3584 CC, Utrecht, The Netherlands
\and
Max-Planck-Institut f\"ur Astrophysik, D-85740 Garching, Germany}

\offprints { N. Langer (email: {\tt N.Langer@astro.uu.nl})}

\date{Received  ; accepted , }

\abstract{
  We present evolutionary calculations for 74 close binaries systems
  with initial primary masses in the range 12...25$\mso$, and initial
  secondary masses between 6 and 24$\mso$.  The initial periods were
  chosen such that mass overflow starts during the core hydrogen
  burning phase of the primary (Case~A), or shortly thereafter
  (Case~B).  We use a newly developed binary code with up-to-date
  physics input. Of particular relevance is the use of OPAL opacities,
  and the time-dependent treatment of semiconvective and thermohaline
  mixing. We assume conservative evolution for contact-free systems,
  i.e., no mass or angular momentum loss from those system except due
  to stellar winds.\\
  We investigate the borderline between contact-free evolution and
  contact, as a function of the initial system parameters.  The
  fraction of the parameter space where binaries may evolve while
  avoiding contact --- which we found already small for the least
  massive systems considered --- becomes even smaller for larger
  initial primary masses. At the upper end of the considered mass
  range, no contact-free Case~B systems exist.  While for primary
  masses of 16$\mso$ and higher the Case~A systems dominate the
  contact-free range, at primary masses of 12$\mso$ contact-free
  systems are more frequent for Case~B. We identify the drop of the
  exponent $x$ in the main sequence mass-luminosity relation of the
  form $L\propto M^x$ as the main cause for this behaviour.\\
  For systems which evolve into contact, we find that this can occur
  for distinctively different reasons. While Case~A systems are prone
  to contact due to reverse mass transfer during or after the
  primary's main sequence phase, all systems obtain contact for
  initial mass ratios below~$\sim 0.65$, with a merger as the likely
  outcome.  We also investigate the effect of the treatment of
  convection, and found it relevant for contact and supernova order in
  Case~A systems, particularly for the highest considered masses.\\
  For Case~B systems we find contact for initial periods above $\sim
  10\,$d. However, in that case (and for not too large periods)
  contact occurs only after the mass ratio has been reversed, due to
  the increased fraction of the donor's convective envelope.  As most
  of the mass transfer occurs conservatively before contact is
  established, this {\it delayed contact} is estimated to yield to the
  ejection of only a fraction of the donor star's envelope.  Our
  models yield the value of $\beta$, i.e., the fraction of the
  primaries envelope which is accreted by the secondary.\\
  We derive the observable properties of our systems after the major
  mass transfer event, where the mass gainer is a main sequence or
  supergiant O~or early~B type star, and the mass loser is a helium
  star. We point out that the assumption of conservative evolution for
  contact-free systems could be tested by finding helium star
  companions to O~stars. Those are also predicted by non-conservative
  models, but with different periods and mass ratios. We describe
  strategies for increasing the probability to find helium star
  companions in observational search programs.\\
\keywords{stars: evolution -- stars: binaries -- stars: blue stragglers -- stars:
  circumstellar matter}
}

\maketitle

\section{Introduction}

To understand the evolution of massive close binary system is relevant
for many current astrophysical problems. Massive binaries provide the
unique means to obtain accurate fundamental stellar properties of
massive stars for various evolutionary stages. I.e., they have been
used to derive masses of Wolf-Rayet stars (van der Hucht 2000) and to
establish mass-luminosity relations for massive stars (Martin et al
1998, Ostrov et al. 2000).  They also lead to the existence of stellar
configurations which do not evolve from single stars and which have
highly exciting observational manifestations.  Examples are the
peculiar supernova 1987A, which several authors attribute to binary
evolution (Podsiadlowski 1992, de Loore \& Vanbeveren 1992, Braun \&
Langer 1995), short-period Wolf-Rayet binaries (e.g., Harries \&
Hilditch 1997), massive X-ray binaries (Chevalier \& Ilovaisky 1998),
supernovae of type~Ib and~Ic (Podsiadlowski et al. 1992), and
$\gamma$-ray burster (Fryer et al. 1999)

However, the evolutionary theory of massive close binaries still
suffers from fundamental uncertainties.  The evolution of such systems
has been modelled by various authors, e.g., Paczy\'nski (1967),
Kippenhahn (1969), de Loore \& De Greve (1992), Pols (1994),
Vanbeveren et al. (1998abc), Wellstein \& Langer (1999). One
longstanding important question in binary models which include mass
transfer is: how much of the transfered matter can actually be
accreted by the secondary star? In most calculations so far, the
fraction $\beta$ of the accreted matter which is retained on the
accreting star is chosen as a constant, with $\beta =0.5$ being a
frequent choice (e.g., Moers \& van den Heuvel 1989, De Greve \&\ De
Loore 1992, De Loore \&\ De Greve 1992, Vanbeveren et al. 1998abc).
This means that half of the matter lost by the donor star is accreted
by the mass gainer; the other half supposed to leave the binary system
due to an undefined force, and carrying an amount of angular momentum
which needs to be described by a second parameter (cf. Huang \& Taam
1990, Podsiadlowski et al. 1992).  Also the assumption of a fully
conservative mass transfer where all matter is accreted by the
secondary, i.e. $\beta = 1$, has been used.  Physical models which
yield the parameter $\beta$ as a function of the system parameters, or
even as a function of time for a give system, are lacking.

Arguments of $\beta \simeq 0.5$ {\em on average} are being used in the
literature (De Greve \&\ de Loore 1992, de Loore \&\ De Greve 1992,
Vanbeveren et al. 1998abc), and correspondingly massive close binary
models using $\beta =0.5$ have been computed (see above).  Here, we
pursue a different approach.  We adopt $\beta = 1$ as long as the two
stars in the binary system do not evolve into contact.  We want to
find out how many massive binaries actually avoid contact when $\beta
= 1$ is used, and whether the result can be compatible with an average
value of $\beta \simeq 0.5$.

Mass transfer in systems with initial mass ratios very different from
unity is known to lead to the rapid expansion of the accreting star
and thus to a contact system (Benson 1970, Ulrich \&\ Burger 1976).
Also systems with large initial periods are prone to develop contact,
as convective envelopes of the donor stars lead to very high mass
transfer rates (cf. Podsiadlowski et al. 1992). Pols (1994) found that
Case~A\footnote{Case~A: mass transfer during central hydrogen burning;
  Case~B: mass transfer after the end of central hydrogen burning;
  Case~C: mass transfer after the end of central helium burning}
systems with primary masses in the range 8...16$\mso$ avoid contact
for initial mass ratios $q\simgr 0.7$ {\em and} initial periods
$\mathrm{P_0} > 1.6\,$d. However, Pols used opacities from Cox \&\ 
Stewart (1970), the Schwarzschild criterion for convection, and
instantaneous thermohaline mixing.  In the present work, we reconsider
the work done by Pols using updated physics input, and we extend the
considered parameters space towards higher masses and larger initial
periods.
 
We describe our computational method in Section~2, and illustrate our
binary model computation at the example of two representative cases in
Section~3.  In Section~4, we present the results of 74 model systems
and discuss the formation of contact as a function of the main system
parameters and the applied stellar physics.  We discuss our results in
Section~5, attempt a comparison with observations in Section~6, and
summarise our main conclusions in Section~7.

\section{Computational methods}

We computed the evolution of massive close binary systems using a
computer code generated by Braun (1997) on the basis of an implicit
hydrodynamic stellar evolution code for single stars (cf. Langer 1991,
1998). It invokes the simultaneous evolution of the two stellar
components of a binary and computes mass transfer within the Roche
approximation (Kopal 1978).

Mass loss from the Roche lobe filling component through the first
Lagrangian point is computed according to Ritter (1988) as
\begin{equation}
  \mathrm{\dot{M}} = \dot{\mathrm{M}}_0 \exp{\frac{\mathrm{R} -
      \mathrm{R_L}}{\mathrm{H_P}}} \ \ \ \ \
      \mathrm{with}\ \ \ \ \ \dot{\mathrm{M}}_0 =
      \frac{1}{\sqrt{e}} \rho v_s \mathrm{Q}.
\end{equation}
$\mathrm{H_P}$ is the photospheric pressure scale height, $\rho$ the
density, $v_s$ the velocity of sound and $Q$ the effective
cross-section of the stream through the first Lagrange pint according
to Meyer \&\ Meyer-Hofmeister (1983).  The distance between the first
Lagrange point and the center of mass which which enter $\mathrm{H_P}$
and $Q$ are interpolated from the tables of Kopal (1978) and Mochnacki
(1984).  We note that this treatment is only correct for stars with
small photospheric pressure scale compared to their radius (Pastetter
\& Ritter 1989). Otherwise, models of Paczy\'nski and Sienkiewicz
(1972), Plavec et al. (1973), Savonije (1978) or Kolb \&\ Ritter
(1990) might be used.  However, this is not required for the binaries
considered in this work.

Due to the exponential dependence of the mass transfer rate on the
stellar radius, an explicit scheme for computing the mass transfer
rate can easily lead to numerical instabilities. Thus, we derive the
mass transfer rate implicitly using an iterative scheme (combined
Secant/Bisection-method; cf. Press et al. 1988), which is numerically
more stable (Braun 1997). However, for each iteration a complete
stellar model needs to be computed.

The change of the orbital period due to the mass transfer and stellar
wind mass loss is computed according to Podsiadlowski et al. (1992),
with the specific angular momentum of the stellar wind material being
determined according to Brookshaw \& Tavani (1993).  The spin angular
momentum of both stars is neglected.  Mass transfer is treated quasi
conservatively. The only mass loss from the system is due to the
stellar winds of both components.  The adopted stellar wind mass loss
rates are described in Wellstein \&\ Langer (1999).

We include time-dependent thermohaline mixing --- i.e. mixing in
radiatively stable regions which occurs due to an outwards increase of
the mean molecular weight. This mixing process is important in stars
which accrete helium-enriched material.  A time-dependent treatment of
thermohaline mixing is important for identifying the borderline
between contact and contact-free close binary systems as it occurs on
the thermal time scale of the accreting star, which --- at the
borderline of contact-free systems --- is just equal to the mass
transfer time scale.  I.e., a time dependent treatment is required to
reproduce the radius evolution of the accreting star during the mass
transfer phase correctly as soon as helium-enriched material is being
transferred (cf., Braun 1997).

Numerically, we treat thermohaline mixing through a diffusion scheme
(Braun 1997). The corresponding diffusion coefficient is based on the
work of Stern (1960), Ulrich (1972), and Kippenhahn et al. (1980); it
reads
\begin{equation}
  D_{thm} = -\alpha_{thm} \frac{3K}{2c_P\rho}
  \frac{\frac{\phi}{\delta}\nabla_\mu}{\nabla_{ad} - \nabla}  ~~,
\end{equation}
where $K=4ac/(3\kappa\rho )$, $\phi =(\partial \ln \rho / \partial \ln
\mu )_{P,T}$, $\delta =-(\partial \ln \rho / \partial \ln T )_{P,\mu
  }$, $\nabla_\mu = d\ln\mu / d\ln P$, $\nabla_{ad}= (\partial \ln T/
\partial \ln P )_{ad}$, and $\nabla = d\ln T / d\ln P$.  The quantity
$\alpha_{thm}$ is a efficiency parameter of order unity, and we use
$\alpha_{thm}=1$ throughout this paper.

\begin{figure*}[ht]
\begin{centering}
\epsfxsize=0.8\hsize
\epsffile{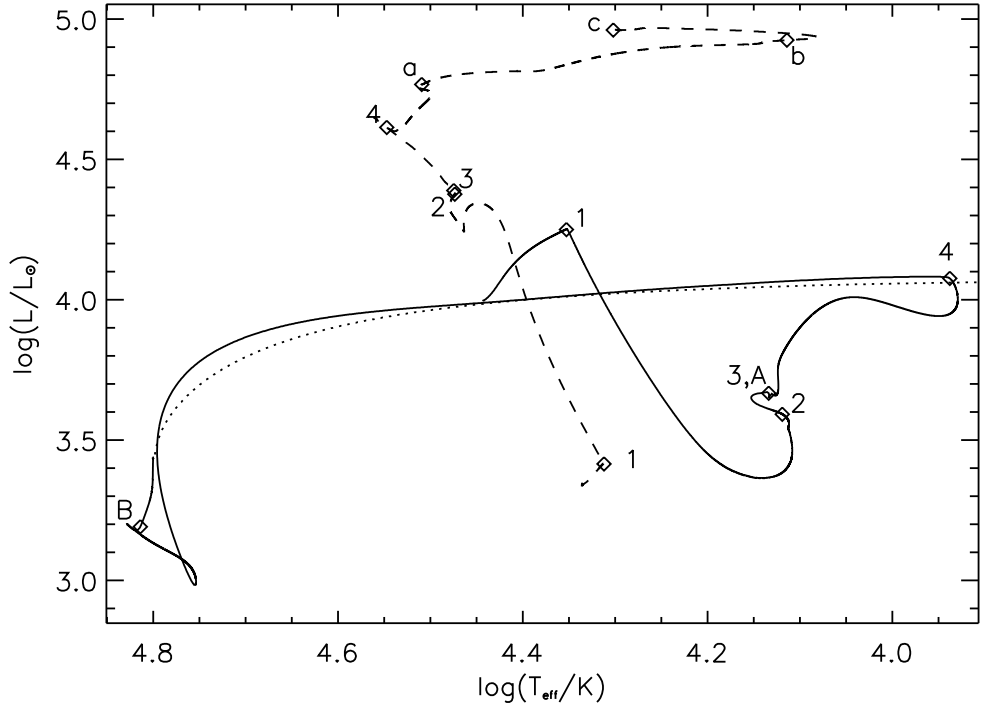}
\caption{Evolutionary tracks of the primary
  (solid and dotted line) and secondary star (dashed line) of our
  case~A binary system No.~31 (initial masses are $12\mso$ and
  $7.5\mso$, the initial period is 2.5$\,$d; cf.~Table~3 in Section~4)
  in the HR~diagram.  Beginning and the end of the mass transfer
  phases are marked with numbers; 1: begin of Case~A, 2: end of
  Case~A, 3: begin of Case~AB, 4: end of Case~AB.  The labels A/a
  designate the end of central hydrogen burning of the
  primary/secondary, B/b the end of central helium burning of the
  primary/secondary, and c the point of the supernova explosion of the
  secondary. In this system, the secondary star ends its evolution
  first. The time of its supernova explosion marks the end of the
  solid line in the track of the primary. The further evolution of the
  primary is shown as dotted line. During this phase, it is treated as
  a single star since the system is likely broken up due to the
  secondary's explosion.}\label{fig:hrda}
\end{centering}
\end{figure*}

The entropy of the accreted material is assumed to be equal to that of
the surface of the mass gainer and the gravitational energy release
due to mass transfer is treated as in Neo et al. (1977).  Convection
and semiconvection are treated as described in Langer (1991) and Braun
\& Langer (1995) (c.f. also Langer et al.  1983).  A semiconvective
efficiency parameter of $\alpha_\mathrm{sc} = 0.01$ is applied in most
models.  We do not include so called convective core overshooting in
our models, as recent evidence implies that the physical effect of
increased convective core masses in massive main sequence stars may be
due to effects of rapid rotation rather than convection (Maeder 1987,
Langer 1992, Heger et al. 2000). As the rotation rates in close
binaries may be reduced due to orbit circularisation and spin-orbit
synchronisation (e.g., Savonije \& Papaloizou 1997), it is unclear to
what extent any convective core mass increase would occur in those
objects.  Opacities are taken from Iglesias \& Rogers (1996).  Changes
in the chemical composition are computed using a nuclear network
including the pp-chains, the CNO-tri-cycle, and the major helium-,
carbon, neon and oxygen burning reactions.  Further details about the
computer program and input physics can be found in Langer (1998) and
Wellstein \& Langer (1999). For all models, a metallicity of 2\% is
adopted.

\section{Case~A and~B evolution: examples}

We have computed 74~Case~A and Case~B systems to derive the
limits of contact free evolution for massive close binaries
within the assumptions made in Section~2.
Before we present the results of these models, we want to discuss
the evolution of one typical Case~A system (No.~31; cf. Sect.~4 below),
and one Case~B system (No.~37) in the following, to illustrate the
capabilities of our method and the type and quality of the derived
results. Those two systems are picked to show one detailed 
example per considered 
initial primary star mass --- i.e., 12 $\Msun$ and 16 $\Msun$; we have
discussed details of models with 25 $\Msun$ initial primary mass 
in Wellstein \& Langer (1999). Furthermore, the two examples
shown in detail below correspond to systems which evolve 
contact-free. Therefore, we could follow the evolution 
of both binary components until the end of carbon burning
(where we stopped the calculations). 

\subsection{Case~A}

\begin{figure}[t!]
\epsfxsize=\hsize
\epsffile{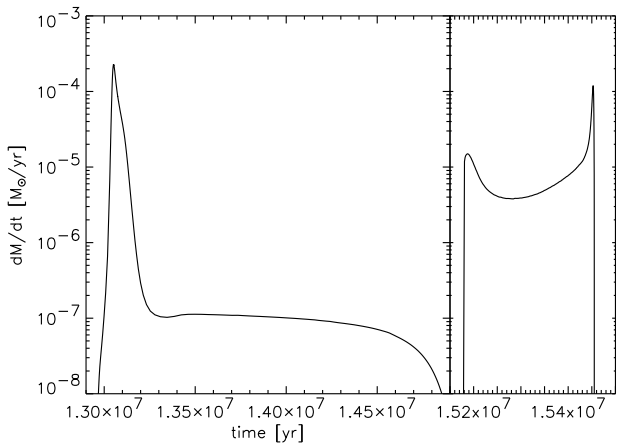}
\caption{Mass transfer rate as function of time for our 
  Case~A system No.31 (initial masses are $12\mso$ and $7.5\mso$, the
  initial period is 2.5$\,$d; cf.~Table~3 in Section~4) The left panel
  shows the rapid and slow Case~A mass transfer, the right panel shows
  the ensuing Case~AB mass transfer.}\label{fig:aroche}
\end{figure}

Table~\ref{fig:taba} and Fig.~\ref{fig:hrda} describe the evolution of
our Case~A system No.~31 (c.f. Table~\ref{fig:systems} below), which
initially consists of a 12 $\Msun$ primary 
\footnote{
  We designate the initially more massive star in the system as the {\em
  primary}, also during its later evolution when it has become the
  less massive and/or less luminous star. The other star is designated
  as the secondary.}
and a 7.5 $\Msun$ secondary
star in a 2.5$\,$d orbit.  Mass transfer starts at a system age of
$\sim 1.3\, 10^7\,$yr, at a core helium mass fraction 
of the primary of~0.89. The mass transfer rate rises to some
$10^{-4}\msoy$ during the so called
rapid Case~A mass transfer phase (Fig.~\ref{fig:aroche}), i.e. to values of
the order of $M_1/\tau_{\rm KH,1}$, $M_1$ being the primary mass and
$\tau_{\rm KH,1}$ the thermal time scale of the primary. The reason is
that the orbital separation of both stars decreases until both
components have the same mass, here about 9.75$\mso$.  
During the slow Case~A mass transfer phase, the
mass transfer is driven by the nuclear evolution of the primary, and the
mass transfer rate drops to some $10^{-7}\msoy$, i.e.  values of the
order of $M_1/\tau_{\rm nuc,1}$.

It is remarkable that the mass transfer does not immediately relax at
the time of minimum orbital separation. In fact,
Fig.~~\ref{fig:aroche} shows that only a small amount of mass ($\sim
0.15\mso$) is transfered during that slow Case~A mass transfer, i.e.
that its rapid phase does not stop until the primary mass is as small
as 4$\mso$. The reason is that, in contrast to the rule of thumb
that stars with radiative envelopes shrink upon mass loss,
the evolved core hydrogen burning 12$\mso$ star has a negative
mass-radius exponent, i.e. its thermal equilibrium radius increases
for decreasing mass as a response to the increasing 
core-mass/envelope-mass ratio. The thermally unstable mass transfer
during the rapid Case~A is ended only when the mass ratio becomes
sufficiently small (cf. Ritter 1988).

During the rapid phase of the Case~A mass transfer the primary becomes
underluminous and the secondary becomes overluminous, due to the rapid
mass loss and gain, respectively.  I.e., they deviate by up to a
factor of two from their thermal equilibrium luminosity (see the kinks
in the evolutionary tracks towards the end of the Case~A mass transfer
in Fig.~\ref{fig:hrda}).  During the slow phase of the Case~A mass
transfer, primary and secondary relax to thermal equilibrium. Due to
the change in total mass, the secondary becomes more luminous than the
primary, even though the primary, after the mass transfer, is strongly
overluminous, i.e. its luminosity exceeds that of a 4$\mso$ single
star by a factor of~10.

\begin{table}[]
\caption{Evolutionary parameters of the Case~A system No.31. 
  The calculation ends after core carbon exhaustion of the primary
  (note that the secondary star produces the first supernova in the
  system).
  The different columns have the following meaning: evolutionary stage
  (ZAMS = zero age main sequence; ECHB = end core hydrogen burning,
  ECHeB = end core helium burning; ECCB = end core carbon burning),
  evolutionary age (ZAMS: $t=0$), primary and secondary star
  mass, orbital period, primary and secondary star orbital velocity.
}
\label{fig:taba}
\begin{center}
\begin{tabular}{l|r r r r r r}
 \hline
~ & time & M$_1$ & M$_2$ & P & $v_{\rm 1}$ & $v_{\rm 2}$ \\
 ~ & Myr & $\Msun$ & $\Msun$ & d & km/s & km/s \\
\hline
ZAMS              & 0     & 12.0 & 7.50 & 2.5  & 162 & 260 \\
begin Case~A      & 12.96 & 11.9 & 7.49 & 2.52 & 162 & 258 \\
end Case~A        & 14.93 & 3.92 & 15.5 & 7.99 & 228 & 57.8\\
ECHB primary      & 15.17 & 3.92 & 15.5 & 7.99 & 228 & 57.8\\
begin Case~AB     & 15.17 & 3.92 & 15.5 & 7.99 & 228 & 57.8\\
end Case~AB       & 15.46 & 1.45 & 17.8 & 102.6& 113 &  9.2\\
ECHB secondary    & 18.94 & 1.45 & 17.7 & 104.2& 112 &  9.2\\
ECHeB primary     & 20.55 & 1.45 & 17.4 & 108.0& 110 &  9.2\\
ECHeB secondary   & 20.82 & 1.45 & 17.2 & 108.1& 109 &  9.2\\
ECCB secondary    & 21.16 & 1.45 & 17.0 & 108.4& 109 &  9.2\\
ECCB primary      & 22.14 & 1.45 & ---  & ---  & --- &  ---\\
 \hline
\end{tabular}
\end{center}
\end{table}

\begin{figure*}[ht!]
\begin{centering}
\epsfxsize=0.80\hsize
\epsffile{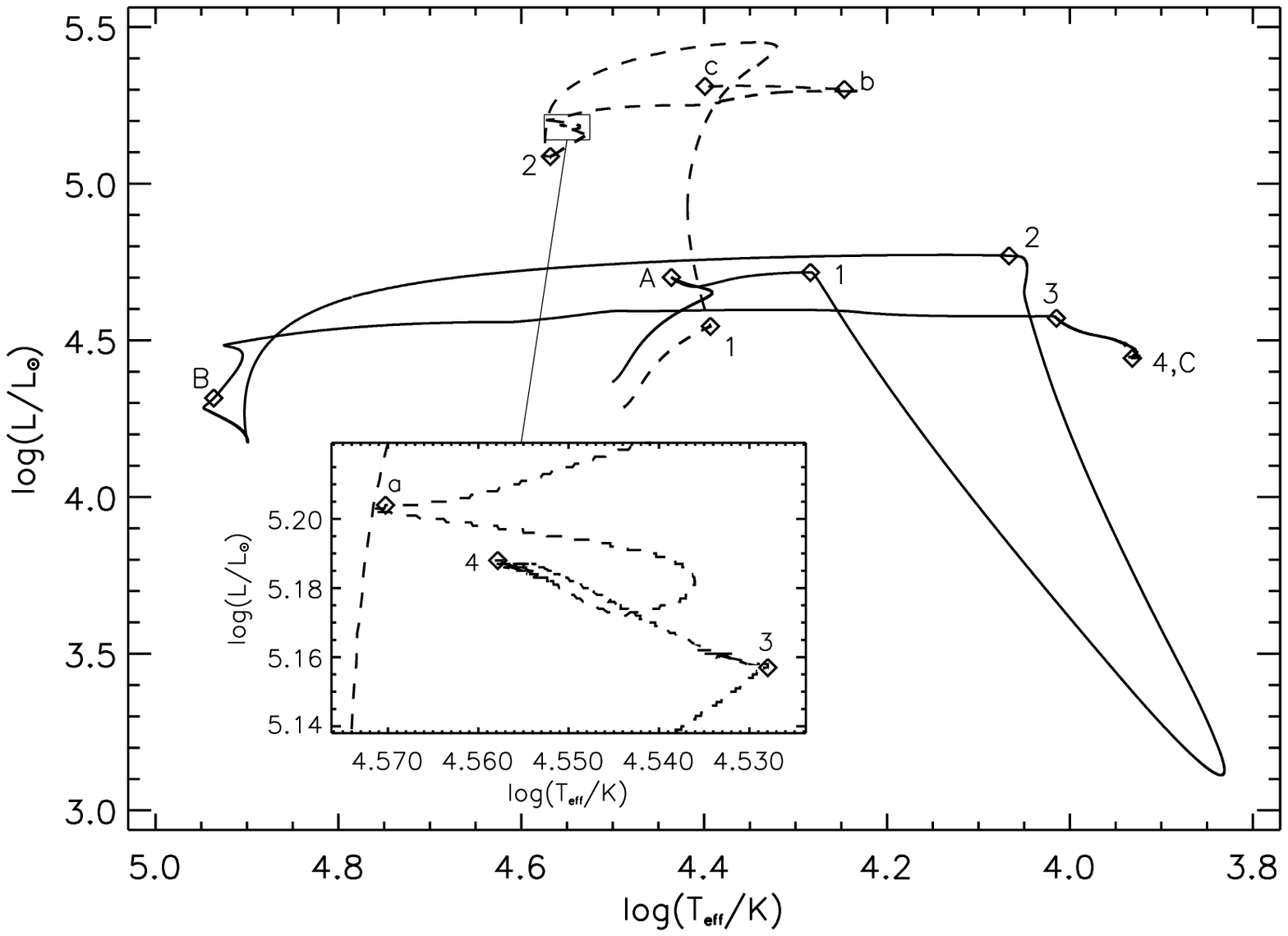}
\caption{Evolutionary tracks of the primary
  (solid and dotted line) and secondary star (dashed line) of our
  case~B binary system No.~37 (initial masses are $16\mso$ and
  $15\mso$, the initial period is 8$\,$d; cf.~Table~3 in Section~4) in
  the HR~diagram.  Beginning and end of the mass transfer phases are
  marked with numbers; 1: beginning of case~B, 2: end of case~B, 3:
  beginning of case~BB, 4: end of case~BB. The labels A/a designate
  the end of central hydrogen burning of the primary/secondary, B/b
  the end of central helium burning of the primary/secondary, and C/c
  the point of their supernova explosion.  }\label{fig:hrdb}
\end{centering}
\end{figure*}

At core hydrogen exhaustion, the primary mass is about 3.9$\mso$.  As
it still contains a hydrogen-rich envelope, a hydrogen burning shell
ignites and the star attempts to expand to red giant dimensions. This
initiates the so called Case~AB mass transfer, which proceeds on the
thermal time scale of the primary and produces correspondingly high
mass transfer rates (cf. Fig.~2). It leads to the loss of almost all
remaining hydrogen and to the extinction of the hydrogen burning
shell. The Case~AB mass transfer reduces the primary mass to about
$1.45\mso$, which corresponds to the mass of the convective hydrogen
burning core at central hydrogen exhaustion.  Due to the extreme mass
ratio (cf. Table~1) the orbital period grows from 8~d to about 100~d
during the Case~AB mass transfer.  The primary spends about 5~Myr for
core helium burning, during which it remains on the helium-main
sequence in the HR~diagram (cf. Fig.~1). The ensuing helium shell
burning expands the star to red giant dimensions, and a so called
Case~ABB mass transfer would occur (Kippenhahn \& Thomas 1979) --- it
does occur in other systems of our sample --- if the secondary star
would not have exploded in the meantime (see below).  We follow the
primary's evolution until core carbon exhaustion. Due to its small
final mass, it is unlikely to produce a supernova explosion. Instead,
mass loss during its red giant stage at moderate rates would suffice
to substantially reduce the total mass before the core mass can grow
to ignite neon burning (Habets 1986ab, Woosley et al. 1995). I.e., our
primary star of initially 12$\mso$ ends as an isolated ONeMg-white
dwarf.

The secondary star grows to 17.81$\mso$ and finishes core hydrogen
burning about 3.5~Myr after the end of the case~AB mass transfer. At
this time it begins to expand to become a supergiant. However, due to
the presence of a strong mean molecular weight barrier above the
convective core at the onset of the mass transfer, it does not
rejuvenate during core hydrogen burning, i.e. it does not adapt its
convective core mass to its new total mass (cf. Hellings 1983, 1984;
Braun \& Langer 1995). Therefore, the secondary star has, after the
mass transfer, a much smaller helium core to envelope mass ratio as a
single star of the same mass. For this reason, it remains in the blue
part of the HR~diagram and does never become a red supergiant (Braun
\& Langer 1995).

Whether or not a secondary star rejuvenates depends strongly on the
--- still poorly known --- efficiency of semiconvection mixing (Langer
et al. 1983, Braun \& Langer 1995).  Were rejuvenation occurring in
our secondary, it would, after the mass transfer, evolve similar to a
single star of $\sim 18\mso$ and attempt to become a red supergiant
after core hydrogen exhaustion. This would lead to a reverse Case~B
mass transfer in this system, with the consequence of contact
evolution and a likely merging of both stars, due to the extreme mass
ratio (cf. System No.~25 below, and Section 5.2, for more details).

Because the secondary star has a small helium core mass, its fuel
supply for core helium burning is small. However, due to its
comparatively large total mass, its luminosity is large.  Both factors
together imply a very short core helium burning life time of only
2~Myr. As a consequence, the secondary overtakes the primary in the
evolutionary status and ends its evolution first.  It explodes as a
nitrogen-rich blue supergiant with properties similar to those of the
progenitor of SN~1987A.  It is helium- and nitrogen enriched due to
the accretion, by factors 1.26 and 5.4, respectively, while carbon and
oxygen are depleted by factors of 0.23 and 0.81 (cf. Wellstein \&
Langer 2001). Its core still has a memory of the lower initial mass:
the helium core mass is only $\sim 4\mso$ at core carbon exhaustion.
This model applied to SN~1987A would predict the
existence of a $1.45\mso$ helium star with a luminosity of $\sim
10^3\lso$ and a temperature of $\sim~55\, 000\,$K in the supernova
remnant. 

In many of our Case~A systems --- always for those with a 12$\mso$
primary --- the secondary is the first star which reaches the
supernova stage.  This supernova reversal is discussed in more detail
in Section~5.2.

\subsection{Case~B}

\begin{figure}[t!]
\epsfxsize=\hsize
\epsffile{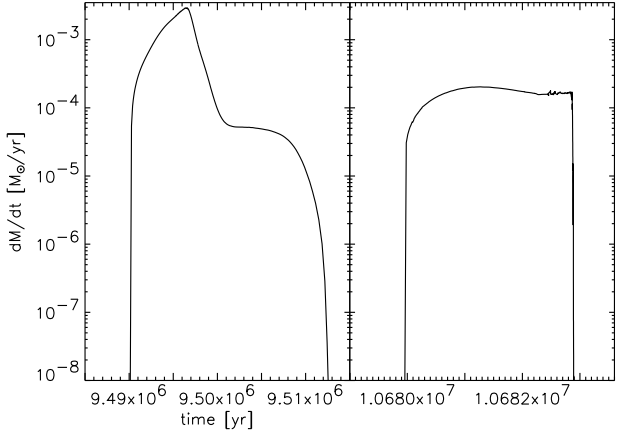}
\caption
{Mass transfer rate as function of time for our Case~B system No.37
  (initial masses are $16\mso$ and $15\mso$, the initial period is
  8$\,$d; cf.~Table~3 in Section~4) The left panel shows the Case~B
  mass transfer, the right panel shows the ensuing Case~BB mass
  transfer.}
 \label{fig:broche}
\end{figure}

Table~\ref{fig:tabb} and Fig.~\ref{fig:hrdb} describe the evolution
of our case~B system No.~37 (cf. Table~\ref{fig:systems}), starting
with a 16$\mso$ primary and a 15$\mso$ secondary in an 8~d orbit.  As
the primary expands on a thermal time scale after core hydrogen
exhaustion, the Case~B mass transfer occurs on the thermal time scale
of the primary, with mass transfer rates of the order of
$M_1/\tau_{\rm KH,1}$ (cf. Sect.~3.1). Both stars are far from thermal
equilibrium during the whole mass transfer phase.  However, also in
this case, a fast and a somewhat slower Case~B mass transfer phase can
be distinguished (Fig.~\ref{fig:broche}), since the orbit shrinks at
first and expands after the mass ratio is reversed.

In the considered system, the mass transfer achieves about $2\,
10^{-3}\msoy$ during the Case~B mass overflow.  This implies an
accretion time scale of the secondary of the same order as its
Kelvin-Helmholtz time scale.  Consequently, the secondary swells
considerably during the Case~B mass transfer. 
While here, contact is not quite achieved, for an initial
period of 9~d instead of 8~d the primary`s radius at the onset of mass
transfer is slightly larger, the mass transfer rate as well, and the
system does evolve into contact at this stage (cf. system No.~39 in
Sect.~4 below).

\begin{table}[h!]
\caption{Evolutionary parameters of the Case~B system No.37.
Both stars are followed beyond core carbon exhaustion.
  The different columns have the following meaning: evolutionary stage
  (ZAMS = zero age main sequence; ECHB = end core hydrogen burning,
  ECHeB = end core helium burning; ECCB = end core carbon burning),
  evolutionary age (ZAMS: $t=0$), primary and secondary star
  mass, orbital period, primary and secondary star orbital velocity.
  }\label{fig:tabb}
\begin{center}
\begin{tabular}{l|r r r r r r}
\hline
~ & time & M$_1$ & M$_2$ & P & $v_{\rm 1}$ & $v_{\rm 2}$ \\
 ~ & Myr & $\Msun$ & $\Msun$ & d & km/s & km/s \\
\hline
ZAMS           & 0     & 16.0 & 15.0 & 8    & 162 &  173\\
ECHB primary   & 9.472 & 15.8 & 14.8 & 8.19 & 160 &  170\\
begin Case~B   & 9.490 & 15.8 & 14.8 & 8.19 & 160 &  170\\
end Case~B     & 9.513 & 3.83 & 26.7 & 94.9 & 128 & 18.3\\
ECHeB primary  & 10.46 & 3.61 & 26.5 & 99.1 & 126 & 17.1\\
begin Case~BB  & 10.68 & 2.81 & 26.5 &102.9 & 126 & 13.4\\
ECCB primary   & 10.68 & 2.33 & 26.8 &170.7 & 109 &  9.4\\
end Case~BB    & 10.68 & 2.32 & 26.8 &172.1 & 108 &  9.4\\
ECHB secondary & 10.95 & ---  & 26.7 & ---  & --- &  ---\\
ECHeB secondary& 11.71 & ---  & 26.0 & ---  & --- &  ---\\
ECCB secondary & 11.83 & ---  & 25.9 & ---  & --- &  ---\\
\hline
\end{tabular}
\end{center}
\end{table}

After the Case~B mass transfer, very little hydrogen is left in the
primary star. It evolves quickly into a hot and compact helium star of
$\sim 3.8\mso$, as which it burns helium in its core for about 1~Myr.
Thereafter, it expands again (cf. Sect.~3.1) and fills its Roche lobe
a second time, which initiates the so called case~BB mass transfer.
During this phase, it loses the major fraction of its helium envelope
--- 0.49 $\Msun$ --- to the secondary.  This mass transfer occurs
again on the thermal time scale of the primary (c.f.
Fig.~\ref{fig:broche}).  Between core carbon exhaustion and neon
ignition, the mass transfer ends.  During the further evolution the
binary system remains detached, but the primary remains an extended
supergiant and nearly fills its Roche lobe. In the final state, the
once 16$\mso$ primary is a 2.3$\mso$ helium giant with a remaining
helium mass of 0.74 $\mso$. The core of the primary will collapse and
produce a supernova explosion. The explosion characteristics may
resemble those of Type~Ib or~Ic supernovae (Shigeyama et al. 1990,
Woosley et al. 1995).  However, the helium envelope of our model star
is very extended; its radius is $\sim 80\rso$.  Theoretical models for
the explosions for such cool helium giants are still missing in the
literature.  Due to the large final mass ratio of 27/2.3, this system
has a good chance to remain bound after the explosion of the primary
component.

The secondary star grows to almost 27$\mso$ during the Case~B mass
transfer. It continues its main sequence evolution thereafter and is
still burning hydrogen in its core when the primary
explodes.  During the Case~BB mass transfer, it accretes 0.49$\mso$ of
almost pure helium from the primary. During this phase, it becomes
very helium and nitrogen rich at the surface. Even though during and
after the Case~BB mass transfer this enrichment is strongly diluted by
thermohaline mixing, the final helium and nitrogen enrichment factors
are 1.25 and~4, i.e. the final surface mass fractions are 0.35 and 0.0041
for helium and nitrogen, respectively.

After core hydrogen exhaustion, the secondary expands, but it remains
a blue supergiant up to its supernova explosion (Fig.~3) due to the
lack of rejuvenation (cf. Sect.~3.1).  The explosion of the secondary
disrupts the binary system with a high probability.

\section{Formation of contact}

In this section, we present results for 74 computed systems.  Rather
than providing a detailed description of the evolution of each system,
we focus on the general trends of key properties as function of the
major initial parameters. Table~3 provides the most important
quantities for all computed systems.  We focus on the various
mechanisms which can drive massive close binary systems into contact.

\begin{table*}[tp!]
\caption{Characteristic quantities for all computed systems:
  system number, initial values of primary mass M$_1$, secondary
  mass M$_2$, and period P$_\mathrm{i}$, and the mass transfer case. 
  Bold values for the initial masses identify the component which
  ends its evolution first.
  Further columns are: the kind of contact experienced by the system
  (no contact='--', premature contact='M', reverse contact='R',
  q-contact='Q', and delayed contact='D'; a lower
  case letter is used if the secondary fills less than 1.5 times its
  Roche lobe; systems No. 52 and 59 experience a second contact
  after a weak q-contact), 
  the core helium mass fraction of the primary at the onset of
  the first mass transfer $Y_{\rm 1}$, and $M_{\rm 1}^{\prime}$, 
  $M_{\rm 2}^{\prime}$, $L_{\rm 1}^{\prime}$, $L_{\rm 2}^{\prime}$,
  R$^\prime_1$, R$^\prime_2$, 
  P$^{\prime}$ denote primary and secondary mass, luminosity,
  effective temperature, radius and
  orbital period after the Case~A/AB or Case~B mass transfer, at a time
  when the core helium mass fraction of the primary has decreased to
  0.8 due to helium burning. 
  $\dot{M}_\mathrm{max}$ is the maximum mass transfer rate (lower
  limits are give for contact systems, values in italic are uncertain) and
  $\Delta M$ is the amount of mass transfered by the time the
  maximum mass transfer rate or contact is reached.
  $\beta$ designates the fraction of the transfered matter
  that is accreted, assuming no further accretion after
  the secondary overfills its Roche lobe more than 1.5 times.
  }\label{fig:systems}
\begin{tabular}{l|l l l l|l l l l l l l l l c c c}
\hline 
No. & M$_1$ & M$_2$ & P$_\mathrm{i}$ & case &  & Y$_\mathrm{1}$ & M$^\prime_\mathrm{1}$ & M$^\prime_\mathrm{2}$ & L$^\prime_\mathrm{1}$ & L$^\prime_\mathrm{2}$ & R$^\prime_1$ & R$^\prime_2$  & P$^\prime$ & $\dot{M}_\mathrm{max}$ & $\Delta M$ & $\beta$ \\
 ~ & \multicolumn{2}{c}{$\Msun$} & d & ~ & ~ & \% & \multicolumn{2}{c}{$\Msun$} & \multicolumn{2}{c}{log($\Lsun$)} & \multicolumn{2}{c}{$\rso$} & d & $\frac{10^{-4} \mso}{\mathrm{yr}}$ & $\mso$\\
\hline
1     & 12 & {\bf 11.5} & 2.5 & A & --& 87 & 1.42 & 21.5 & 2.98 & 4.99 & 0.3 & 16.5 & 218 & {\it 0.71} & 1.8 & 1.00\\
2     & {\bf 12} & 11   & 3   & B & --& 98 & 2.44 & 20.4 & 3.93 & 4.74 & 0.8 & 7.2  & 54  &       4.0  & 7.3 & 1.00\\
3     & {\bf 12} & 11   & 6   & B & --& 98 & 2.38 & 20.4 & 3.82 & 4.79 & 0.6 & 6.8  & 115 &       10   & 7.3 & 1.00\\
4     & {\bf 12} & 11   & 10  & B & --& 98 & ---  & ---  & ---  & ---  & --- & ---  & --- &       17   & 7.5 & 1.00\\
5     & {\bf 12} & 11   & 16  & B & --& 98 & ---  & ---  & ---  & ---  & --- & ---  & --- &       25   & 7.5 & 1.00\\
6     & {\bf 12} & 11   & 22  & B & --& 98 & ---  & ---  & ---  & ---  & --- & ---  & --- &       28   & 7.6 & 1.00\\
7     & 12 & 11         & 30  & B & D & 98 & ---  & ---  & ---  & ---  & --- & ---  & --- &    $>$33   & 6.5 & 0.68\\
8     & 12 & 11         & 40  & B & D & 98 & ---  & ---  & ---  & ---  & --- & ---  & --- &    $>$33   & 5.7 & 0.60\\
9     & 12 & 10.5       & 2   & A & M & 75 & ---  & ---  & ---  & ---  & --- & ---  & --- & {\it 0.67} & 1.5 & 0.70\\
10    & 12 & {\bf 10.5} & 2.5 & A & --& 87 & 1.42 & 20.7 & 2.97 & 4.87 & 0.3 & 7.9  & 190 & {\it 0.93} & 2.0 & 1.00\\
11    & 12 & 10         & 2   & A & M & 76 & ---  & ---  & ---  & ---  & --- & ---  & --- &       0.82 & 1.6 & 0.68\\
12    & {\bf 12} & 10   & 16  & B & --& 98 & 2.43 & 19.4 & 3.81 & 4.71 & 0.6 & 6.6  & 255 &       22   & 7.0 & 1.00\\
13    & 12 & 10         & 22  & B & D & 98 & ---  & ---  & ---  & ---  & --- & ---  & --- &    $>$28   & 6.9 & 0.73\\
14    & 12 & {\bf 9.5}  & 2   & A & --& 76 & 1.29 & 19.4 & 2.93 & 5.00 & 0.4 &32.3$^e$& 171&      1.8  & 2.0 & 1.00\\
15    & 12 & {\bf 9.5}  & 2.5 & A & --& 88 & 1.41 & 19.7 & 2.96 & 4.79 & 0.3 & 7.5  & 166 & {\it 1.2}  & 2.0 & 1.00\\
16    & {\bf 12} & 9.5  & 3.5 & B & --& 98 & 2.45 & 18.9 & 3.94 & 4.63 & 0.8 & 6.3  & 50  & {\it 4.0}  & 2.9 & 1.00\\
17$^a$& 12 & {\bf 9}    & 2   & A & R & 76 & 1.15 & 19.4 & 2.67 & 4.87 & 0.3 & 9.0  & 214 & {\it 1.2}  & 1.7 & 1.00\\
18    & {\bf 12} & 9    & 3.5 & B & --& 98 & 2.32 & 18.5 & 3.74 & 4.63 & 0.6 & 6.5  & 53  &       6.5  & 3.4 & 1.00\\
19    & {\bf 12} & 9    & 6   & B & --& 98 & 2.36 & 18.5 & 3.81 & 4.63 & 0.6 & 6.2  & 87  &       9.9  & 7.4 & 1.00\\
20    & {\bf 12} & 9    & 10  & B & --& 98 & ---  & ---  & ---  & ---  &  ---& ---  & --- &       15   & 7.0 & 1.00\\
21    & {\bf 12} & 9    & 16  & B & d & 98 & ---  & ---  & ---  & ---  & --- & ---  & --- &       28   & 7.0 & 1.00\\
22    & {\bf 12} & 9    & 22  & B & D & 98 & ---  & ---  & ---  & ---  & --- & ---  & --- &    $>$29   & 5.9 & 0.62\\
23    & 12 & {\bf 8.5}  & 2.5 & A & --& 88 & 1.43 & 18.8 & 2.97 & 4.72 & 0.3 & 6.8  & 135 & {\it 1.6}  & 2.1 & 1.00\\
24    & 12 & {\bf 8}    & 2   & A & --& 77 & 1.12 & 18.5 & 2.64 & 4.84 & 0.3 & 8.5  & 189 &       3.0  & 2.3 & 1.00\\
25$^c$& 12 & 8          & 2.5 & A & R & 87 & 1.66 & 17.9 & 3.18 & 4.68 & 0.4 & 6.5  & 81.5&       3.6  & 2.5 & 1.00\\
26    & {\bf 12} & 8    & 3.5 & B & q & 98 & ---  & ---  & ---  & ---  & --- & ---  & --- &       7.3  & 3.0 & 1.00\\
27    & {\bf 12} & 8    & 6   & B & --& 98 & 2.37 & 17.5 & 3.79 & 4.55 & 0.6 & 5.8  & 72  &       8.9  & 3.5 & 1.00\\
28    & {\bf 12} & 8    & 10  & B & --& 98 & ---  & ---  & ---  & ---  & --- & ---  & --- &       15   & 6.8 & 1.00\\
29    & {\bf 12} & 8    & 16  & B & D & 98 & ---  & ---  & ---  & ---  & --- & ---  & --- &    $>$24   & 6.0 & 0.63\\
30    & 12 & 7.5        & 2   & A & q & 78 & ---  & ---  & ---  & ---  & --- & ---  & --- &       2.1  & 2.0 & 1.00\\
31    & 12 & {\bf 7.5}  & 2.5 & A & --& 89 & 1.45 & 17.8 & 2.99 & 4.64 & 0.3 & 6.3  & 104 &       2.5  & 2.0 & 1.00\\
32    & 12 & 7.5        & 3.5 & B & Q & 98 & ---  & ---  & ---  & ---  & --- &  --- & --- &    $>$8.1  & 2.8 & 0.27\\
33    & 12 & 6.5        & 2   & A & Q & 78 & ---  & ---  & ---  & ---  & --- &  --- & --- &    $>$4.6  & 1.4 & 0.13\\
 ~    &  ~ &  ~         &  ~  & ~ & ~  &  ~ & ~    &  ~   &  ~   &  ~   & ~   & ~    &  ~ &      ~     & ~   &     \\
34    & {\bf 16} & 15.7 & 3.2 & A & --& 91 & 2.79 & 27.8 & 3.88 & 5.14 & 0.5 & 9.3  & 75  &       3.5  & 3.6 & 1.00\\
35    & {\bf 16} & 15.7 & 6   & B & --& 98 & 3.49 & 27.5 & 4.17 & 5.14 & 0.6 & 8.3  & 86  &       22   & 7.7 & 1.00\\
36    & 16 & 15         & 2.5 & A & M & 80 & ---  & ---  & ---  & ---  & --- & ---  & --- &       2.5  & 2.6 & 0.70\\
37    & {\bf 16} & 15   & 8   & B & --& 98 & 3.63 & 26.7 & 4.20 & 5.10 & 0.7 & 8.9  & 96  &       29   & 8.0 & 1.00\\
38    & {\bf 16} & 15   & 9   & B & d & 98 & 3.64 & 26.7 & 4.22 & 5.10 & 0.7 & 8.1  & 107 &       32   & 8.1 & 1.00\\
39    & 16 & 15         & 15  & B & D & 98 & ---  & ---  & ---  & ---  & --- & ---  & --- &    $>$37   & 6.7 & 0.48\\
40    & 16 & 14         & 2   & A & M & 69 & ---  & ---  & ---  & ---  & --- & ---  & --- &       2.2  & 2.2 & 0.86\\
41$^a$& 16 & {\bf 14}   & 2   & A & --& 68 & 2.12 & 27.0 & 3.68 & 5.21 & 0.6 & 11.0 & 110 & {\it 1.2}  & 2.1 & 1.00\\
42    & 16 & {\bf 14}   & 2.5 & A & --& 80 & 2.39 & 26.5 & 3.78 & 5.17 & 0.6 & 11.0 & 97.5&       3.1  & 2.9 & 1.00\\
43    & 16 & 14         & 3   & A & --& 87 & 2.52 & 26.6 & 3.75 & 5.10 & 0.5 & 9.0  & 101 &       2.0  & 2.8 & 1.00\\
44    & {\bf 16} & 14   & 6   & B & --& 98 & 3.60 & 25.8 & 4.19 & 5.04 & 0.6 & 7.8  & 67  &       24   & 7.9 & 1.00\\
45    & {\bf 16} & 14   & 9   & B & D & 98 & 3.64 & 25.7 & 4.21 & 5.04 & 0.7 & 7.6  & 98  &       31   & 8.2 & 0.64$^f$\\
\hline
\multicolumn{11}{l}{continued on next page}\\
\hline
\end{tabular}
\end{table*}
\begin{table*}[tp!]
\begin{tabular}{l|l l l l|l l l l l l l l l c c c}
\hline
\multicolumn{11}{l}{continued from previous page}\\
\hline
No. & M$_1$ & M$_2$ & P$_\mathrm{i}$ & case &  & Y$_\mathrm{1}$ & M$^\prime_\mathrm{1}$ & M$^\prime_\mathrm{2}$ & L$^\prime_\mathrm{1}$ & L$^\prime_\mathrm{2}$ & R$^\prime_1$ & R$^\prime_2$  & P$^\prime$ & $\dot{M}_\mathrm{m}$ & $\Delta M$ & $\beta$ \\
 ~ & \multicolumn{2}{c}{$\Msun$} & d & ~ & ~ & \% & \multicolumn{2}{c}{$\Msun$} & \multicolumn{2}{c}{log($\Lsun$)} & \multicolumn{2}{c}{$\rso$} & d & $\frac{10^{-4} \mso}{\mathrm{yr}}$ & $\mso$\\
\hline
46    & 16 & 13         & 2   & A & M & 69 & ---  & ---  & ---  & ---  & --- & ---  & --- &       3.0 & 2.5 & 0.68\\
47    & 16 & {\bf 13}   & 2.5 & A & --& 80 & 2.28 & 26.1 & 3.65 & 5.11 & 0.5 & 9.1  & 99  & {\it 2.1} & 2.4 & 1.00\\
48$^c$& {\bf 16} & 13   & 2.5 & A & --& 79 & 2.47 & 25.6 & 3.77 & 5.10 & 0.6 & 7.6  & 78  &       4.3 & 3.0 & 1.00\\
49    & {\bf 16} & 13   & 3   & A & --& 88 & 2.64 & 25.3 & 3.84 & 5.07 & 0.5 & 9.1  & 64  &       5.2 & 3.5 & 1.00\\
50$^c$& {\bf 16} & 13   & 3   & A & --& 87 & 2.66 & 25.2 & 3.81 & 5.08 & 0.5 & 7.8  & 62,5&       5.1 & 3.2 & 1.00\\
51    & {\bf 16} & 13   & 4   & B & d & 98 & 3.55 & 24.8 & 4.19 & 4.99 & 0.7 & 8.2  & 41  &       18  & 6.9 & 1.00\\
52    & {\bf 16} & 13   & 6   & B &  D& 98 & 3.70 & 24.7 & 4.24 & 4.95 & 0.7 & 6.8  & 57  &       27  & 7.8 & 0.62$^f$\\
53    & 16 & 12         & 1.5 & A & M & 54 & ---  & ---  & ---  & ---  & --- & ---  & --- &       3.5 & 2.4 & 0.71\\
54    & 16 & 12         & 1.7 & A & M & 61 & ---  & ---  & ---  & ---  & --- & ---  & --- &       3.7 & 2.5 & 0.75\\
55    & 16 & {\bf 12}   & 2   & A & --& 70 & 2.07 & 24.8 & 3.57 & 5.17 & 0.5 & 11.8 & 93  &       4.2 & 2.8 & 1.00\\
56    & {\bf 16} & 12   & 4   & B & d & 98 & 3.55 & 23.9 & 4.17 & 4.93 & 0.6 & 7.0  & 37  &       18  & 6.3 & 1.00\\
57    & {\bf 16} & 12   & 6.5 & B & D & 98 & ---  & ---  & ---  & ---  & --- & ---  & --- &       26  & 7.7 & 0.38$^f$\\
58    & 16 & {\bf 11}   & 2   & A & --& 71 & 2.00 & 24.1 & 3.56 & 5.10 & 0.5 & 10.8 & 89  &       5.7 & 3.1 & 1.00\\
59$^c$& 16 & 11         & 2   & A &qR & 69 & 2.10 & 24.0 & 3.54 & 5.13 & 0.5 & 8.4  & 77  &       5.9 & 3.1 & 1.00\\
60    & {\bf 16} & 11   & 3   & A & --& 89 & 2.52 & 23.8 & 3.74 & 4.95 & 0.5 & 7.3  & 61  & {\it 3.8} & 2.8 & 1.00\\
61    & {\bf 16} & 11   & 3.2 & A & --& 92 & 2.81 & 23.3 & 3.87 & 4.93 & 0.5 & 7.4  & 42  &       8.2 & 4.0 & 1.00\\
62    & {\bf 16} & 10   & 2.5 & A & q & 81 & 2.28 & 23.1 & 3.77 & 4.97 & 0.6 & 7.8  & 65  &       8.7 & 3.5 & 1.00\\
63    & {\bf 16} & 10   & 3   & B & q & 89 & ---  & ---  & ---  & ---  & --- & ---  & --- &       9.9 & 3.9 & 1.00\\
64    & 16 & 9          & 2.5 & A & Q & 82 & ---  & ---  & ---  & ---  & --- & ---  & --- &       12  & 3.4 & 0.16$^f$\\
  ~   &  ~ &  ~         &  ~  & ~ & ~ &  ~ & ~    &  ~   &  ~   &  ~   & ~   & ~    &  ~  &  ~        &     &     \\
65    & 25 & {\bf 24}   & 3.5 & A & --& 83 & 5.15 & 40.6 & 4.53 & 5.54 & 0.6 & 13.7 & 41  & {\it 3.0} & 4.6 & 1.00\\
66$^a$& {\bf 25} & 24   & 3.5 & A & --& 83 & 5.21 & 40.6 & 4.54 & 5.54 & 0.6 & 12.4 & 40  & {\it 3.0} & 4.2 & 1.00\\
67$^b$& {\bf 25} & 24   & 3.5 & A & --& 82 &---$^d$& 40.5&---$^d$& 5.54&---$^d$&11.2&---& {\it 3.0} & 4.6 & 1.00\\
68    & 25 & {\bf 24}   & 5   & B & d & 98 & 7.28 & 39.2 & 4.85 & 5.45 & 0.8 & 11.9 & 31  &       30  & 8.2 & 1.00\\
69    & 25 & 24         & 9   & B & D & 98 & ---  & ---  & ---  & ---  & --- & ---  & --- &    $>$37  & 7.4 & 0.43\\
70    & 25 & {\bf 23}   & 4   & A & --& 87 & 5.31 & 39.4 & 4.56 & 5.49 & 0.6 & 12.4 & 39  & {\it 4.0} & 5.2 & 1.00\\
71    & 25 & {\bf 22}   & 2.5 & A & --& 70 & 4.80 & 39.6 & 4.45 & 5.55 & 0.7 & 14.2 & 39  & {\it 2.7} & 2.9 & 1.00\\
72$^c$& {\bf 25} & 22   & 2.5 & A & --& 68 & 4.68 & 38.9 & 4.44 & 5.62 & 0.7 & 10.5 & 39  &       5.3 & 3.6 & 1.00\\
73    & {\bf 25} & 19   & 4   & A & --& 88 & 5.26 & 35.9 & 4.55 & 5.38 & 0.7 & 8.7  & 30  &       6.3 & 4.9 & 1.00\\
74    & {\bf 25} & 16   & 4   & A & --& 89 & 5.22 & 33.3 & 4.54 & 5.29 & 0.6 & 8.8  & 24  &       8.8 & 5.0 & 1.00\\
\hline
\end{tabular}

$^a$ $\alpha_\mathrm{sc} = 0.02$\\
$^b$ $\alpha_\mathrm{sc} = 0.04$\\
$^c$ Schwarzschild criterion instead of Ledoux criterion used\\
$^d$ Only secondary is computed but primary should not be very different from No. 65 and 66\\ 
$^e$ Secondary at ECHeB\\
$^f$ Treated as if $\beta$ = 1\\
\end{table*}

\subsection{Contact formation due to small or large periods}

\subsubsection{Case~B}

Case~B systems with an initial mass ratio close to~1 and a rather
short initial period ({\em early} Case~B) are known to have a good
chance to avoid contact during the Case~B mass transfer (cf. Pols
1994).  Our Table~3 contains 15~examples, i.e. systems No.~2...6, 12,
16, 18...20, 27, 28 (initial primary mass of $12\mso$), and No.~35, 37
and 44 (initial primary mass of $16\mso$). The fact that {\em no} such
system with an initial primary mass of $25\mso$ could be found is
discussed in Sect.~5.1.  It is often assumed that {\em late} Case~B
systems, i.e. such with a relatively large initial period, evolve
through a common envelope stage (Podsiadlowski et al. 1992, Vanbeveren
1998abc).  In the following, we will have a closer look to the
transition region in between both extremes.

\begin{figure}[t]
\epsfxsize=\hsize
\epsffile{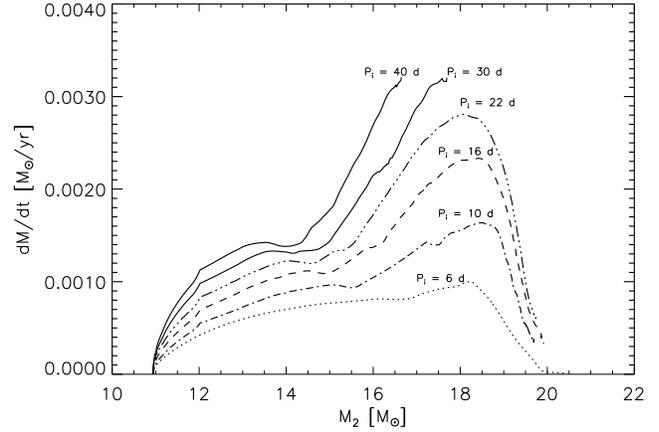}
\caption{Mass transfer rates as a function of secondary mass
  for Case~B systems initially consisting of a 12$\mso$ and a 11$\mso$
  star with initial periods in the range 6...40$\,$d (No.~3,4,5,6,7
  and 8).  The systems with 30 and 40 days initial period form
  contact, at which time we stop the calculation.}\label{fig:m12bXXrl}
\end{figure}

Fig.~\ref{fig:m12bXXrl} shows the mass transfer rates as a function
of the secondary mass for Case~B binaries with initial primary and
secondary masses of 12$\mso$ and 11$\mso$, respectively, but different
initial periods in the range 6...40$\,$d (Systems No.~3 to~8).
Generally, initially wider systems develop larger mass transfer rates.
However, it is interesting to consider the time dependence of the mass
transfer rate of these systems in some detail.  The first maximum in
the mass transfer rate is related to the end of decrease of the
orbital separation at mass ratio one.  I.e., when the primary has
become the less massive component in the system, the period and the
Roche lobe of the primary begin to grow, and the increase of the mass
transfer rate flattens.  This happens with some time delay due to the
finite thermal response time of the primary.

Later during the mass transfer, the orbit widens significantly.
Therefore, the primary becomes more extended and its surface
temperature decreases. In the wider systems considered in
Fig.~\ref{fig:m12bXXrl}, the outer envelope of the primary becomes
even convectively unstable. It is the drop of the adiabatic
mass-radius exponent $\zeta_{\rm ad}$ (cf. Ritter 1988) which leads to
another rise of the mass transfer rate even though the orbit widens.
We designate this feature as {\it delayed contact}.  Table~3 shows nine
systems which evolve delayed contact.  It is restricted to the Case~B
and has no counterpart in our Case~A systems (cf. Sect.4.1.2).

\begin{figure}[t]
\epsfxsize=\hsize
\epsffile{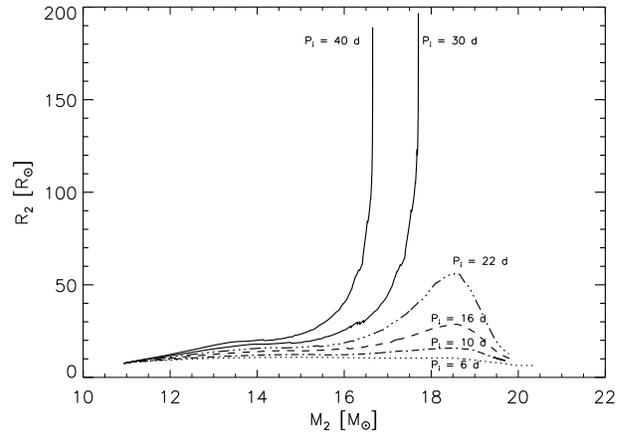}
\caption{Radius of the secondary during mass transfer as a function
  of its mass, for the same systems as in Fig.~\ref{fig:m12bXXrl}. In
  the systems with 30 and 40 days initial period the large increase in
  radius leads to contact, at which time we stop the calculation.
  }\label{fig:m12bXXr}
\end{figure}

Fig.~\ref{fig:m12bXXr} displays the radius evolution of the
secondary stars of the systems considered in Fig.~\ref{fig:m12bXXrl}.
Initially, the radii do not grow significantly, although it can be
seen seen that larger radii are obtained for larger initial periods,
i.e. for larger mass transfer rates. Later on, mass transfer rates in
excess of $\sim 3\, 10^{-3}\msoy$ make the secondary swell enormously.
This mass transfer rate corresponds roughly to $\dot M = R L / (G M)$,
i.e. to the condition $M/ \dot M = G M^2 / (R L)$ for the secondary.
In the two systems with initial periods of 30 and 40~days, the radius
increases until the secondary also fills its Roche lobe and a contact
system is formed. This effect marks the limit of contact free systems
towards larger periods.

It is remarkable that approximately 5~$\Msun$ for the 40~d binary
(system No.~8) and 6~$\Msun$ for the 30~d binary (system No.~7) are
transfered before the contact occurs (cf. Fig.~6).  From
Fig.~\ref{fig:m12bXXr} we see that contact occurs the earlier (in
terms of already transfered mass) the larger the initial system period
is. I.e., there is a continuous transition towards very wide binaries
with convective primaries, which evolve into contact at the very
beginning of the mass transfer process (Podsiadlowski et al. 1992).

According to the estimate of Webbink (1984), which compares the
orbital energy of the binary with the binding energy of the envelope,
our System No.~7 does not merge during the Case~B mass transfer.
Instead, with an efficiency parameter of $\alpha_\mathrm{ce}=1$, it
obtains a period of about 7.5~d in a common envelope evolution by
expelling the remaining envelope of the primary of $\sim 3\mso$ from
the system.

In summary, in delayed contact systems a conservative part of Case~B
mass transfer is followed by a non-conservative one, leading to an
oscillation of the orbit: it first shrinks until the mass ratio is~1,
then widens up to the end of the conservative part, and then shrinks
again due to mass and angular momentum loss from the system. For our
System No.~7, the parameter~$\beta$ for the complete Case~B mass
transfer, i.e., the ratio of the amount of mass accreted by the
secondary to the amount lost by the primary, is about $\beta \simeq
6\mso / 9\mso \simeq 0.67$.

After the delayed contact and common envelope evolution, System No.~7
consists of a $\sim 20\mso$ main sequence star --- the secondary ---
and a $\sim 2.4\mso$ helium star in a 7.5~d orbit.  The further
evolution (which we have not computed) depends on the 
post-common envelope period. For 7.5~d,
the primary fills its Roche lobe as it expands to a helium giant, and transfers
part of its helium envelope to the secondary in a Case BB mass
transfer, similar to that discussed in Sect.~3.2 for system No.~37.

In contrast, the systems which avoid contact during the Case~B mass
transfer all together can often, even if the secondary evolves faster,
avoid the reverse Case~B mass transfer if the secondary star does not
rejuvenate (cf. values of $T_{\rm eff,s}^{\prime}$ in Table~3).  After
the Case~B mass transfer, their periods are typically larger than
50~d.

We find that in our most massive systems, with a primary mass of
25$\mso$, even {\rm early} Case~B mass transfer leads into contact,
even for an initial mass ratios very close to~1 (cf. systems No.~68
and~69 in Table~3). In the two considered systems, large amounts of
mass are transferred conservatively before contact occurs (8.2$\mso$
and 7.4$\mso$; cf. Table~3), i.e., we have a delayed contact situation.
In system No.~68, the contact is marginal, in system No.~69 it is not.

\subsubsection{Case~A}

\begin{figure}[t]
\epsfxsize=\hsize
\epsffile{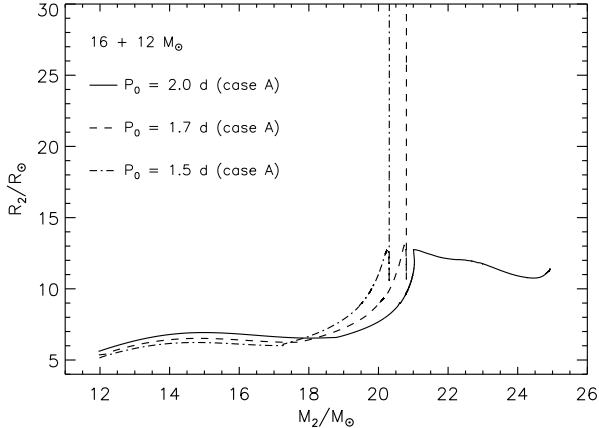}
\caption{Secondary radius as a function of its mass during Case~A mass 
  transfer, for three systems starting out with a 16$\mso$ and a
  12$\mso$ component, with initial periods in the range 1.5...2~d
  (Systems No. 53...55).  In the two closer systems, the secondaries
  terminate core hydrogen burning during the slow phase of the Case~A
  mass transfer. Their ensuing rapid expansion leads to contact. In
  the 2~d period system No.~55 (solid line), Case~AB mass transfer
  starts (at $M_2\simeq 21\mso$) before the secondary has exhausted
  hydrogen in the core, and contact is avoided (see text).
  }\label{fig:m16p12R}
\end{figure}

Fig.~\ref{fig:m16p12R} demonstrates a completely different mechanism
of contact formation operating in Case~A systems, which occurs for
{\em short} periods (cf. also Pols 1994). It shows the radius
evolution of three secondaries in Case~A systems which have identical
initial parameters except the period (systems No.~53...55 in
Table~\ref{fig:systems}). All three secondaries evolve through the
rapid part of the Case~A mass transfer (cf. Sect.~3.1) without
expanding significantly (which is different for initial mass ratios $q
\simle 0.65$; cf. Sect.~5.1). However, in the two closer systems, the
mass transfer appears so early in the evolution of the primary that
the now more massive secondary finishes core hydrogen burning first.
It then starts to expand, but soon fills its Roche lobe and attempts a
reverse Case~B mass transfer (in systems No.~53 and~54). Thus, both
stars now fill their Roche lobes and come into contact while the
primary is still burning hydrogen in its core; we designate this as
{\em premature contact}.  The likely outcome of premature contact is a
merger because the corresponding periods are too small for a
successful common envelope ejection. This effect marks the lower
period limit for contact free binaries.

In model No.~55, the primary ends core H-burning first.  Therefore,
Case~AB mass transfer (cf. Section~3.1) sets in before the secondary
expands strongly. This has two effects: it drives the two stars
further apart --- the period increases from 5.24~d at the beginning to
93~d at the end of Case~AB mass transfer ---, and it enhances the
mixing of hydrogen into the core (cf. Braun \& Langer 1995) and thus
delays the evolution of the secondary. I.e., if the initial period
exceeds a critical value, Case~AB mass transfer starts before the
secondary expands and premature contact can be avoided. If the
secondary star does not rejuvenate, reverse Case~B mass transfer is
avoided as well. Curiously, the secondary star in our Case~A system
No.~1 finishes core hydrogen burning just after the end of the Case~A
mass transfer and before the beginning of the Case~AB mass transfer.
However, the accretion due to the Case~AB mass overflow leads to a
shrinkage of its radius, and contact is avoided (see also Fig.~11
below).

The lower period limit for contact free binaries depends sensitively
on the convection criterion. As mentioned in Section~2, we compute
stellar models using the Ledoux criterion for convection and
semiconvection with an efficiency parameter of $\alpha_{\mathrm{sc}} =
0.01$.  The semiconvective mixing efficiency --- which is infinite in
the case of the Schwarzschild criterion for convection --- determines
to a large extent whether an accreting core hydrogen burning star
rejuvenates or not (Braun \& Langer 1995). If it fails, which is the
more likely for less efficient semiconvection and the more advanced
the star is in its evolution when it starts to accrete, it retains an
unusual structure, with a smaller helium core mass and a larger
hydrogen envelope mass compared to single stars. As such a structure
keeps the star relatively compact during all the advanced burning
stages --- i.e., it avoids the red supergiant stage (Braun \& Langer
1995) --- it will not fill its Roche lobe after core hydrogen burning.
This is the situation in all our Case~A systems which avoid premature
contact and have initial mass ratios of $q \simgr 0.65$.

For models computed with the Schwarzschild criterion, the secondaries
in Case~A systems rejuvenate during or shortly after the rapid Case~A
mass transfer. This results in longer main sequence life times for the
secondaries, and thus leads to less premature contact situations.
I.e., the critical initial period for premature contact is shifted to
smaller values. In this case, the secondaries all expand to red
supergiants. Assuming the primary star is still present at that time,
this is likely to initiate reverse Case~B mass transfer.  As the life
time of a rejuvenating star is extended, the primaries have often
finished their evolution by that time and the binary may even be
disrupted (cf. Pols 1994).  However, as the life time of a low mass
helium star (i.e. the primary of such a system) can be comparable to
the remaining hydrogen burning life time of the very massive
secondary, reverse Case~B mass transfer onto the low mass helium star
may also occur. In this case, the helium star builds up a hydrogen
envelope and swells to red giant dimensions. We designate the ensuing
contact situation in Table~3 as {\em reverse contact}.

This scenario is illustrated by five of our Case~A systems (No.~25,
48, 50, 59, and~72) which are computed using the Schwarzschild instead
of the Ledoux criterion (marked ``{\it c}'' in Table~3).  No.~25 and
59, initially having a 12$\mso$ and a 16$\mso$ primary, a mass ratio
of 0.67 and 0.69, and a period of 2.5~d and 2.0~d, respectively, avoid
premature contact.  However, they encounter reverse contact when the
primaries have burnt 91\% and 80\% of their helium in the core.  Both
systems avoid contact altogether when they are computed with our
standard convection physics, and evolve a reverse supernova order.
I.e., an increase of the semiconvective mixing efficiency may reduce
the number of systems with premature contact but increase the number
with reverse contact.  From system No.~25 we conclude that, were the
Schwarzschild criterion correct, the majority of Case~A binaries with
12 $\Msun$ primaries would evolve into reverse contact.
Comparing our models No.~58 and~59, which have identical initial
conditions but vary in the treatment of convection, shows that also in
systems with 16 $\Msun$ primaries the lower period limit for contact
free systems increases.

In the three other systems computed with the Schwarzschild criterion,
systems No.~48, 50, and~72, the primaries explode as supernovae before
the reverse mass transfer can occur. Two of these (No.~48 and~72),
when computed with our standard convection physics, have a reverse
supernova order (see systems No.~47 and~71 in Table~3), the other
(No.~50) does not (cf. system No.~49 in Table~3).

Thus, the lower period limit for contact free evolution with the
Schwarzschild criterion is expected between the lower period limit and
the limit for reverse supernova order for Ledoux criterion and
$\alpha_\mathrm{sc} = 0.01$. The one early case~A system containing a
25 $\Msun$ primary (No. 72) also shows no reverse mass transfer before
the supernova explosion of the primary, despite being a system with
reverse supernova order when computed with our standard convection
physics. We conclude that for large masses, the amount of contact free
case~A systems does not change very much for different convection
physics. On the contrary, at 12 $\Msun$ primary mass and below no
contact free case~A system are expected at all if the Schwarzschild
criterion is used.

\subsection{Contact formation due to extreme mass ratios}

\subsubsection{Case~B}

While the previous two subsections concerned contact formation due to
large (delayed contact) or small (premature and reverse contact)
initial periods, we investigate here contact formation due to large or
small initial mass ratios.

\begin{figure}[t!]
\epsfxsize=\hsize
\epsffile{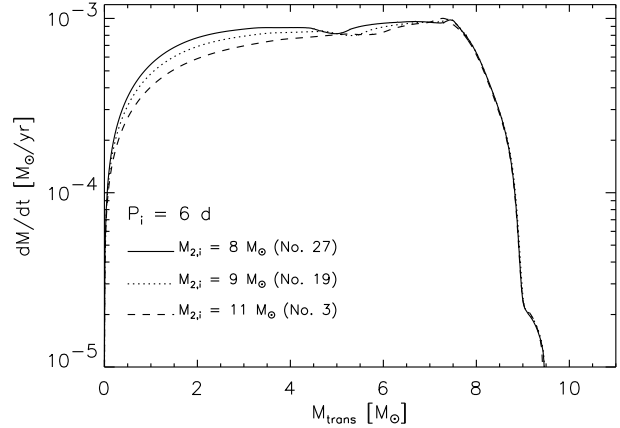}
\caption{Mass transfer rates as function of the transfered amount of
  mass for three Case~B systems with 12$\mso$ primaries.  The initial
  masses of the secondaries are 11$\mso$ (system No.~3), 9$\mso$
  (No.~19), and 8$\mso$ (No.~27), as indicated. The initial period is
  6~d for all three systems.  }\label{fig:m12X9m2}
\end{figure}

Fig. \ref{fig:m12X9m2} shows the mass transfer rate as function of
the transferred amount of mass for three Case~B systems with identical
initial primary mass (12$\mso$) and period (6~d), but different
initial secondary masses. For smaller initial mass ratios $q_{\rm
  i}=M_{2,{\rm i}} / M_{1,{\rm i}}$, the mass transfer rates are
somewhat larger in the early part of the mass transfer, since the
orbit shrinks more rapidly for smaller $q_{\rm i}$.  For conservative
evolution (also ignoring stellar winds), the minimum separation
depends on initial separation $ d_{\rm i}$ and mass ratio $q_{\rm i}$
as $d_{\rm min} = d_{\rm i} \left( 4 q_{\rm i} \over (q_{\rm i} +1)^2
\right)^2$.  As, in a given system, period and orbital separation
scale as $P^2 \propto d^3$, this means that in systems with the same
initial period, the primary star is, at the time when $P=P_{\rm min}$,
squeezed into a smaller volume for smaller $q_{\rm i}$. This makes the
mass transfer rate during the first phase of Case~B mass transfer
larger for smaller $q_{\rm i}$.

However, the maximum mass transfer rate, which is achieved after mass
ratio reversal (cf. Sect.~4.1), is very similar in all three systems,
even though the post-mass transfer periods (115~d, 87~d, and 72~d for
secondary initial masses of 11$\mso$, 9$\mso$, and 8$\mso$,
respectively) are not the same.  Also the amount of mass which is
transferred is independent of the secondary mass --- it consists of
the whole hydrogen-rich envelope of the primary.

\begin{figure}[t!]
\epsfxsize=\hsize
\epsffile{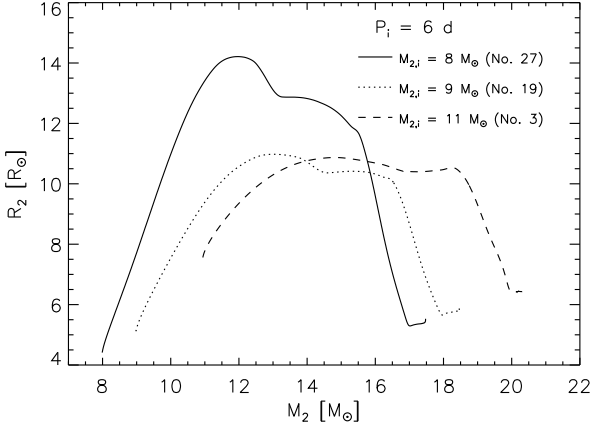}
\caption{Radii of the secondary stars as function of their mass
  for the same systems as shown in Fig.~\ref{fig:m12X9m2}.
  }\label{fig:m12X9m1}
\end{figure}

Whether or not contact is reached is here mostly determined by the
reaction of the secondary star to the accretion.
Fig.~\ref{fig:m12X9m1} shows that the primary with the lowest
initial mass swells most during the mass transfer.  The smallest
initial mass ratio for which contact is avoided corresponds roughly to
the condition that the mass accretion time scale of the secondary
$M_{\rm 2}/\dot M_{\rm 2}$ remains smaller than its thermal time scale
$G M_{\rm 2}^2 / (R_{\rm 2} L_{\rm 2})$.  We designate contact due to
this reason as {\em $q$-contact}.

As $q$-contact is established during the first phase of Case~B mass
transfer, it is quite distinct from delayed contact (Sect.~4.1). As
$q$-contact evolves early during Case~B mass transfer, it does not
allow a major amount of mass to be accreted by the secondary. I.e.,
either the major part of the primary's hydrogen-rich envelope can be
ejected from the system, or both stars merge.  The situation that the
secondary star can accrete significant amounts {\em after} a
non-conservative early $q$-contact phase is unlikely since it would
bring both stars even closer together.

In principle, there is also a maximum initial mass ratio for
contact-free evolution of Case~B systems: for $q_{\rm i}=1$, both
stars age at the same rate and attempt to expand to red giants
simultaneously. However, since the limiting value is very close to
$q_{\rm i}=1$, we ignore this effect here as statistically
insignificant.

\subsubsection{Case~A}

\begin{figure}[t!]
\epsfxsize=\hsize
\epsffile{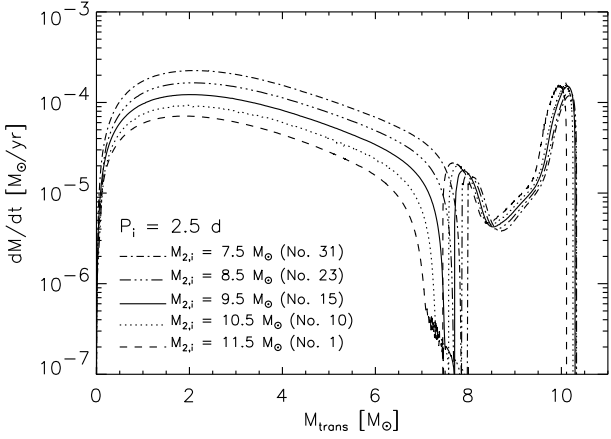}
\caption{Mass transfer rate as a function of transfered amount of mass, 
  for five Case~A systems with the same initial primary mass
  ($12\mso$) and initial period (2.5~d).  The initial secondary masses
  are 11.5$\mso$ (System No.~1), 10.5$\mso$ (No.~10), 9.5$\mso$
  (No.~15), 8.5$\mso$ (No. 23), and 7.5$\mso$ (No.~31).
  }\label{fig:m12X4m2}
\end{figure}

\begin{figure}[t!]
\epsfxsize=\hsize
\epsffile{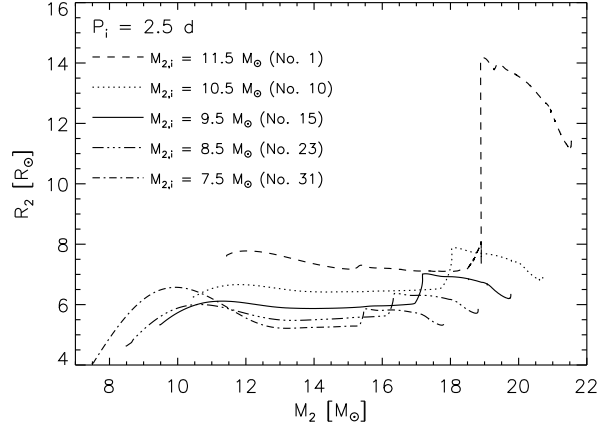}
\caption{Radii of the secondary stars as function of their mass
  for the same systems shown in Fig.~\ref{fig:m12X4m2}.  The first
  maximum is achieved during rapid Case~A mass transfer.  The second
  maximum is achieved during slow Case~A mass transfer due to the
  nuclear evolution of the secondary. The radius decrease after the
  second maximum is due to the Case~AB mass transfer.
  }\label{fig:m12X4m1}
\end{figure}

The maximum mass transfer rate in Case~A systems is achieved during
the rapid mass transfer phase while the orbit is shrinking.
Fig.~\ref{fig:m12X4m2} shows the mass transfer rates of five Case~A
systems with identical initial conditions except for the initial
secondary mass and thus the initial mass ratio $q_{\rm i}$. As shown
in Sect.~4.2.1, the smaller $q_{\rm i}$ the larger is the maximum mass
transfer rate since the primary is squeezed into a smaller volume.
Also, more mass is transfered during the rapid Case~A mass transfer
for smaller $q_{\rm i}$.  As for Case~B, q-contact may occur for small
$q_{\rm i}$ due to higher maximum mass transfer rates {\em and} due to
larger thermal time scales of the secondary, the latter effect being
the more important one. In Fig.~\ref{fig:m12X4m1}, all systems shown
avoid q-contact, but the relative radius increase during the rapid
Case~A mass transfer is clearly larger for the systems with smaller
initial secondary masses.

The second expansion of the secondaries during the Case~A mass
transfer (Fig.~\ref{fig:m12X4m1}) is due to their nuclear evolution
during the slow mass transfer phase.  It is ended by the onset of
Case~AB mass transfer. The secondary star in system No.~1 has already
ended core hydrogen burning and expanded appreciably before the onset
of Case~AB mass transfer.  The onset of Case~AB mass transfer stops
its expansion and prevents the system from evolving into premature
contact.  However, Figs.~\ref{fig:m12X4m2} and~\ref{fig:m12X4m1}
demonstrate, that for too large mass ratios, massive Case~A binaries
evolve into premature contact as in the case of too small initial
periods (cf. Section~4.1.2). I.e., while Case~B systems evolve into
contact only when their initial mass ratio is too large, Case~A
binaries can do so for too large or too small initial mass ratio.

Fig.~\ref{fig:m12X4m2} shows also, that the total amount of mass
which is transfered during case~A and AB is nearly independent of
$q_{\rm i}$.  This can not be expected in general, as we shall see
below that the primary mass after Case~A and AB mass transfer depends
on the core hydrogen abundance at the onset of mass transfer and on
the initial mass ratio (cf. Fig.~\ref{fig:eala} below). While
smaller initial mass ratios lead to smaller primary masses, the larger
Roche lobe due to the lower secondary mass (for the same period) leads
to mass transfer only later in the evolution of the primary and thus
to larger post-mass transfer primary masses. Both effects cancel each
other in the models shown in Fig.~\ref{fig:m12X4m2}.

\section{Discussion}

\subsection{The contact-free regime}

\begin{figure}[t]
\epsfxsize=\hsize
\epsffile{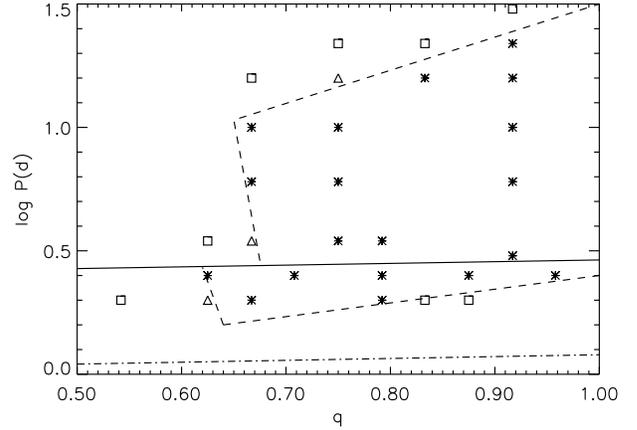}
\caption{Distribution of all computed binaries with 12$\mso$ primaries in the 
  log~P~versus~q-diagram. Asterisks mark contact-free systems, while
  squares mark systems which evolve into contact. Systems marked with
  triangles are borderline cases, i.e., they evolve into a short
  contact phase but the secondary radius never exceeds its Roche
  radius by more than a factor~1.5. The solid line separates Case~A
  (below) and Case~B systems. All case~A systems for this primary mass
  have a reverse supernova order. The dashed lines indicate the
  boundary between contact-free and contact evolution. The
  dashed-dotted line is defined by the condition that the primary
  fills its Roche lobe already on the zero age main sequence.
  }\label{fig:m12pq}
\end{figure}

\begin{figure}[t]
\epsfxsize=\hsize
\epsffile{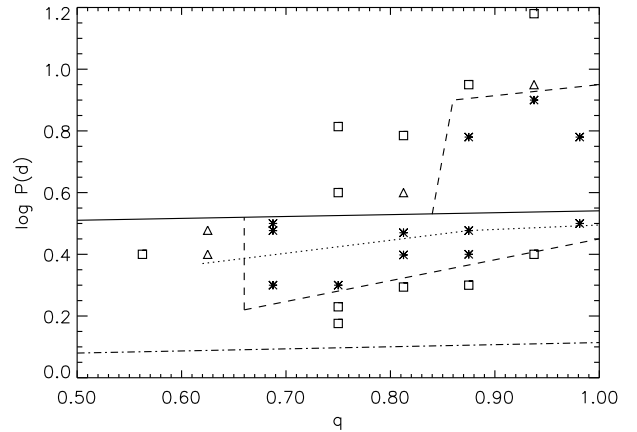}
\caption{The same as in Fig.~\ref{fig:m12pq} but for systems
  with initial primary masses of $16\mso$. The dotted line divides the
  contact-free Case~A systems into such with normal and reverse
  supernova order, the latter occurring in the systems below the line.
  }\label{fig:m16pq}
\end{figure}

Fig.~\ref{fig:m12pq} and~\ref{fig:m16pq} summarise the range of
contact-free evolution for systems with 12$\mso$ and 16$\mso$
primaries, respectively. For all mass transfer cases and primary
masses, there is a rather well defined critical initial mass ratio
$q_{\rm c}$ such that only systems with larger $q_{\rm i}$ avoid
contact. For $q_{\rm i} < q_{\rm c}$, they evolve into $q$-contact.
Likely, most of the systems undergoing $q$-contact merge, since
contact occurs early during the mass transfer, at a small orbital
separation.  We find $q_{\rm c}\simeq 0.65$ for all Case~A systems
with primaries in the initial mass range 12...25$\mso$. For Case~B
systems, $q_{\rm c}$ decreases from $q_{\rm c} \simeq 0.7$ at 12$\mso$
to $q_{\rm c} \simeq 0.8$ at 16$\mso$ and $q_{\rm c} \simeq 1$ at
25$\mso$ initial primary mass.

The range of contact-free Case~A systems is further limited by the
possibility of {\em premature contact}, which occurs for initial
periods below a certain threshold, but also for initial mass ratios
{\em above} a critical value, i.e. too close to one.  We should note
that for even smaller initial periods than investigated here,
premature contact may occur already before the secondary has finished
core hydrogen burning (cf. Pols 1994).  In any case, the outcome of
premature contact is most likely the merging of both stars.  The
possibility of reverse contact and of a reverse supernova order for
the Case~A systems, and their dependence on the semiconvective mixing
time scale are discussed in the next section.

As described in Section 4.1.2, the uncertain efficiency of
semiconvective mixing introduces a considerable uncertainty to the
evolution of Case~A systems. Semiconvective mixing affects the
rejuvenation process in an essential way, as it is this process which
transports hydrogen into the convective core of the accretion stars.
While for infinitely fast semiconvective mixing --- which corresponds
to the use of the Schwarzschild criterion for convection --- all
accretion stars rejuvenate (Hellings 1983), this is not the case for
finite semiconvective mixing time scales (Braun \& Langer 1995).
Longer mixing time scales lead to shorter core hydrogen burning life
times of the secondaries.  Therefore, slow semiconvection --- as
assumed here --- leads to earlier contact. Nevertheless, slow
semiconvection leads perhaps not to a smaller total number of contact
systems, since for those systems which avoid premature contact, the
chance of reverse contact during core helium burning of the secondary
is much smaller than in the case of fast semiconvective mixing (cf.
Sect.~4.1.2).

We note that although semiconvection can be important for the upper
end of the mass spectrum of our study (cf. also Wellstein \& Langer
1999), another effect dominates for the less massive Case~A systems.
In our systems with an initial primary mass of 12$\mso$, the
luminosity ratio after the mass transfer is very large, of the order
of~100, while it is only of the order of~10 for our most massive
systems (see Table~3).  The reason for the drop of the luminosity
ratio for higher primary masses is simply the flattening of the
stellar mass-luminosity relation for increasing mass. As the
luminosity ratio reflects the ratio of the evolutionary speeds of both
components, it is clear that with a post-mass transfer value of~100,
only very late Case~A systems can avoid contact --- independent of
whether the secondary rejuvenates or not.  The large luminosity ratio
for lower primary initial masses is also the main reason why the
initial parameter space for contact-free evolution becomes smaller for
lower mass systems.

Finally, there is an upper period which limits the contact-free regime
of the Case~B systems. Our calculations have shown that systems close
to this limit may experience {\em delayed contact}, with the
consequence of considerable accretion on the secondary, and a high
chance to avoid the merging of both stars. Initially even wider
systems may correspond more to the standard common envelope scenario,
again with the possibility to avoid a merger, but likely without
significant accretion of the secondary.

By comparing Figs.~\ref{fig:m12pq} and~\ref{fig:m16pq}, and including
our results for even more massive systems (see Table~3, and Wellstein
\& Langer 1999), it becomes clear that the range of contact-free
Case~A systems becomes smaller for smaller initial primary masses,
while that for contact-free Case~B systems becomes smaller for larger
ones. E.g., there are {\em no} contact-free Case~B systems for initial
primary masses of 25$\mso$.  We conclude that at lower masses,
conservative evolution is to be expected mainly from close Case~B
systems, while at larger masses Case~A systems dominate the
conservative evolution.  Considering that massive binaries with
initial periods of up to some thousand days may interact, and adopting
an equal number of binaries per log~P bin, up to one third of all
interacting systems in the considered mass range and with
$\mathrm{q_i} \gtrsim 0.7$ may evolve conservatively.

\subsection{Remnant types}

\begin{figure}[t!]
\epsfxsize=\hsize
\epsffile{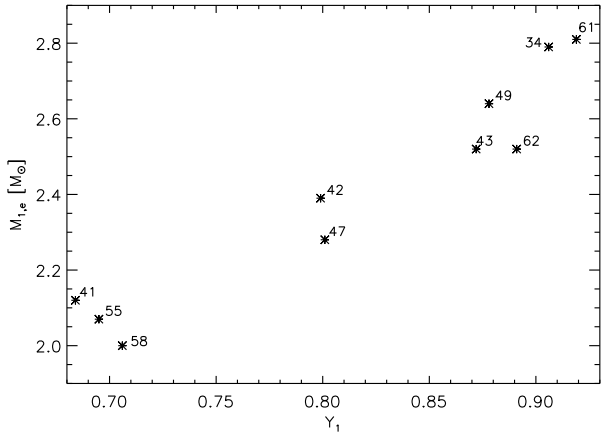}
\caption{Masses after Case~AB mass transfer for
  primaries with an initial mass of $16\mso$, as function of their
  central helium abundance at the onset of Case~A mass transfer.  For
  comparison, the post Case~B mass transfer masses of $16\mso$
  primaries are all about 3.6$\mso$.  }\label{fig:eala}
\end{figure}

Those of our systems which avoid merging evolve into OB+He~star
systems. We discuss in the following the further evolution of the
primaries, the secondaries, and of merger stars.

\subsubsection{Primaries}
Table~3 shows that the helium star masses after the Case~AB/B mass
transfer in our systems fall into the range 1...7$\mso$.  The helium
star masses in the Case~B systems correspond simply to the helium core
mass of the primary at core hydrogen exhaustion, with a very small
scatter ($+0.1\mso$). Fig.~\ref{fig:eala} shows that the situation
is more complex for the Case~A systems.  We find a strong dependence
of the helium star mass on the initial period, i.e. on the central
hydrogen abundance of the primary when the mass transfer begins. We
also find a weaker dependence on the initial mass ratio (cf., systems
No.~41, 55, 58).  Fig.~\ref{fig:eala} demonstrates that our 16$\mso$
Case~A primaries become 2...3$\mso$ helium stars --- compared to a
Case~B post-mass transfer mass of $\sim 3.6\mso$, which is independent
of other parameters. We point out that these strongly reduced Case~A
primary masses are often ignored in simplified binary studies which
rely on single star evolutionary models.

The further evolution of the helium stars depends strongly on their
mass, and partly on the evolution of their companion stars.  The more
massive helium stars in our mass range correspond to Wolf-Rayet stars,
and will undergo wind mass loss before they explode as Type~Ib/Ic
supernova. Those helium stars with masses below $\sim 4\mso$ expand to
giant dimensions after core helium exhaustion. Depending on whether
the companion star is still present at that time, this may lead to
Case~ABB/BB mass transfer, and a further reduction of the mass of the
helium star by $\sim 0.5\mso$ or so (cf. system No.~37 in Sect.~3.2).
Independent of this, the helium stars with final masses in the range
2...4$\mso$ are likely to explode as helium {\em giants}.  The
effect of the large radii of these stars on the supernova light curve
is still unexplored. Helium stars in the mass range 1.4...2$\mso$
could, in principle, also make it to core collapse, but as their
post-core helium burning life time is rather large it may be more
likely that stellar winds will reduce their mass below the
Chandrasekhar mass and they end up as white dwarfs, like their
counterparts with masses between 1...1.4$\mso$.

We point out that all 12$\mso$ primaries in our contact-free Case~A
systems end up as white dwarfs, and even some 16$\mso$ primaries may
do so (e.g., system No.~55). I.e., the limiting initial mass
separating white dwarf formation from core collapse is largely
increased in Case~A binaries, compared to single stars. We note that
this effect exists also for Case~B binaries, strongly so for primaries
which undergo a Case~BB mass transfer, but at a moderate level even
otherwise. The same effect for the limiting initial mass between
neutron star and black hole formation has been discussed by Wellstein
\& Langer (1999).

Finally, we can not exclude that some of the helium stars with masses
close to but above the Chandrasekhar mass develop a degenerate
C/O-core and ignite carbon explosively. In that case, they would
produce a Type~Ia supernova. However, in the 1.45$\mso$ helium star of
system No.~31 (cf. Sect.~3.1), a luminosity of $\sim 10^4\lso$ during
shell helium burning implies a growth rate of the C/O-core mass of
$\sim 10^{-6}\msoy$, which is fast enough to allow carbon to ignite
non-explosively.

\subsubsection{Secondaries}
All our secondary stars evolve to core collapse. In Case~A systems,
about half of the secondaries does so before the primary has ended its
evolution (cf. Table~3). As most of our primaries did not rejuvenate,
they remain blue supergiants throughout their post-main sequence
evolution (cf.  Figs.~1 and~3). When they explode as Type~II
supernova, they will thus resemble SN~1987A. We noted in Sect.~3.1
that the secondary of system No.~31 provides a viable progenitor model
for SN~1987A.  In this case, there should be a $\sim 1.5\mso$ helium
star left in the supernova remnant. 
In similar systems, however, the supernova order 
is not reversed and no such helium star would be expected (cf. Table~3). 

\subsubsection{Merger}

Several of our systems evolve into contact and are likely to lead to a
merger. While it is beyond our scope to predict the properties of the
merged stars, we can distinguish two types of events. If the merging
occurs during Case~A mass transfer and before the secondary has
terminated core hydrogen burning, the merged object will still be a
core hydrogen burning star. Its surface may be enriched in hydrogen
burning products, but otherwise it might resemble a normal main
sequence star in many respects.

However, a merger during a Case~B or reverse Case~B mass transfer (cf.
Sect.~4.1.1; note that the latter may even occur during a Case~A mass
transfer; cf. Fig.~7) has a different result. Since the original
primary has already developed a compact hydrogen-free core, the merger
star will have a hydrogen-free core with a mass close to that of the
primary.  It is uncertain how much of the hydrogen-rich material is
lost during the merging process, and whether some hydrogen-rich matter
can penetrate into the helium core. However, if the mass of the merger
star exceeds the mass of the primary star, which appears likely, then
the helium core mass of the merger star is much smaller than the
helium core mass of a single star of comparable mass.  I.e., its
structure is similar to that of our non-rejuvenating accretion stars
during core helium burning. As a consequence of their small helium
core mass, these stars may avoid to evolve into red supergiants
(Podsiadlowski et al. 1990, 1992).

\subsubsection{Reverse supernova order}

In our Case~A binaries, a reversal of the supernova order appears in
all systems containing a 12 $\Msun$ primary (c.f.
Fig.~\ref{fig:m12pq}) and in some systems containing a 16 $\Msun$
primary (cf. Fig.~\ref{fig:m16pq}).  At 12$\mso$, supernova order
reversal depends not so much on whether or not the secondary star
rejuvenates. Instead, the mass ratio after the rapid phase of the
Case~A transfer, and thereby the luminosity ratio, is essential. Due
to the strong decrease of the primaries mass, it becomes a factor of
about 100 less luminous than the secondary. Therefore, independent of
how much fresh hydrogen is mixed into the secondaries core, it
finishes its evolution first as long as no reverse mass transfer
occurs.

However, in case of rapid semiconvective mixing, the secondary may
rejuvenate, with the consequence of reverse mass transfer {\em before}
it can evolve into a supernova.  Due to the extreme mass ratio, the
reverse mass transfer is expected to be non-conservative. Its outcome
is either a merger or a short period double helium star binary.
Assuming the latter, we can estimate whether or not the supernova
order would be reversed. In system No.~17 --- computed with
$\alpha_\mathrm{sc} = 0.02$ --- the secondary is more evolved at time
of reverse mass transfer, i.e. both stars burn helium in their cores,
with central helium mass fractions of 0.5/0.3 for the
primary/secondary. Thus the supernova order is expected to be
reversed. For system No.~25 computed with Schwarzschild criterion, it
is less clear what happens. The primary has already spent about
$6~10^6$~yr at central helium burning and has reached a core helium
mass fraction of~0.09.  The secondary has reached a core helium mass
fraction of 0.63 after only $3~10^5$ years of central helium burning.
Because the primary's helium shell burning phase is expected to last
more than $1~10^6$ years (cf. system No. 31 in Table~\ref{fig:taba})
it appears possible that also in this system the supernova order is
reversed, if the system does not merge.

For systems with an initial primary mass of 25 $\Msun$ the convection
physic dominates the question whether the supernova order is reversed.
This can be seen in systems No. 65, 66 and 67, where the value for the
efficiency parameter for semiconvective mixing $\alpha_\mathrm{sc}$ is
the only difference. Only in system No.~65 with $\alpha_\mathrm{sc} =
0.01$ the supernova oder is reversed. With $\alpha_\mathrm{sc} \ge
0.02$ no reversal appears. Also the case~B system 68 reverses its
supernova order with $\alpha_\mathrm{sc} = 0.01$. This shows that the
mass of the primary after mass transfer in this systems is not very
important because the nuclear time scales of the stars in this mass
range vary less with mass than for lower mass stars. Also the early
case~A binary No. 71 has a reverse supernova order in contrast to
system No. 72, which has the same parameters, but is calculated with
the Schwarzschild criterion.  The systems with 16 $\Msun$ primary are
an intermediate case, but seems to behave more like the systems
containing a 25 $\Msun$ primary (c.f. systems No. 47 and 48).

Whether or not a reverse mass transfer occurs before the primary
becomes a compact object depends strongly on the semiconvective mixing
speed. For fast semiconvective mixing, the secondaries become red
supergiants. Then, the crucial question is not if reverse mass
transfer occurs but when it occurs. A shorter semiconvective mixing
time scale has two effects. First, the maximum radius of the secondary
after core hydrogen exhaustion increases (cf.  Wellstein \&\ Langer
1999). This makes it more likely for the secondary to fill its Roche
lobe and transfer matter back to the primary. Second, the time the
secondary spends on the main sequence increases, as more hydrogen is
mixed into the convective core during central hydrogen burning. Thus
it is less likely that the reverse mass transfer occurs before the
primary has finished its evolution and has become a compact object.

Although in general, systems which have a normal supernova order would
be able to avoid reverse mass transfer and vice versa, some systems
which show a reverse supernova order for slow semiconvective mixing
($\alpha_\mathrm{sc} = 0.01$) avoid reverse mass transfer during the
lifetime of the primary when the Schwarzschild criterion is used, as
shown by systems No. 48 and 72.

\subsection{Comparison to previous work}

Pols (1994) presented 24~binary models for initial primary masses in
the range 8...16$\mso$, mostly for Case~A systems.  Since he used the
Schwarzschild criterion for convection, the secondary components
rejuvenated in all computed cases.  As a consequence, contact phases
during slow case~A mass transfer occurred for shorter periods than in
our models.  Furthermore, due to the rejuvenation, a supernova
explosion of the secondary as the first star could only appear after a
reverse mass transfer phase. This mass transfer appears at very large
mass ratios and must therefore be non-conservative. As discussed by
Pols, it is an open question whether or not such a binary merges or is
able to expel the secondary's hydrogen-rich envelope to form a close
binary consisting of two helium stars. As the helium star resulting
from the secondary is then the more massive component, a supernova
order reversal is possible if the merging of both stars can be
avoided.

Sybesma (1985) computed 5 binary sequences starting with a primary
mass of 20~$\Msun$, a secondary mass of 10~$\Msun$, and periods
ranging from 1.5 up to 10 days.  His results differ from ours mainly
due his assumption of huge convective core overshooting.  This leads
to much more massive helium-rich cores in the main sequence models.
Consequently, he obtains Case~A mass transfer for initial periods
which fall in our Case~B regime. As Case~A mass transfer is more
likely conservative in this mass range, Sybesma obtains conservative
evolution for larger periods than we (e.g.  10 days), and for smaller
initial mass ratio (q=0.5).

Several authors specify the fraction $\beta$ of the primary's envelope
which is accreted by the secondary in advance. E.g., based on
observations, De~Greve \&\ de~Loore (1992) and de~Loore \&\ De~Greve
(1992) use $\beta =0.5$, which means half of the primary's envelope
mass leaves the system, together with an amount of angular momentum
which needs to be specified additionally. While an average value of
$\beta =0.5$ appears in fact possible, our models favour the idea that
$\beta$ decreases continuously for larger initial periods, with $\beta
= 1$ for systems which avoid contact and $\beta = 0$ for large initial
periods. E.g., we obtained $\beta = 0.68$ and $\beta = 0.60$ for our
systems No.~7 and~8 (cf. Table 3). By comparison with De~Greve \&\ 
de~Loore (1992) and de~Loore \&\ De~Greve (1992), we note that the
contact-free regime is expanded somewhat if $\beta =0.5$ is used ab
initio.

\section{Comparison with observations}

\begin{table*}[tp!]
\caption{Observable quantities for our contact-free systems.
  After system number, initial values of primary mass M$_1$, secondary
  mass M$_2$, and period P$_\mathrm{i}$, we give quantities obtained
  after the Case AB or Case~B mass transfer, at a time
  when the core helium mass fraction of the primary has decreased to
  0.8 during core helium burning: 
  primary and secondary mass and mass ratio,
  bolometric luminosities, effective temperatures (where no wind effects
  are included for the helium stars), stellar radii,
  the radius of the secondary in units of its Roche radius,
  orbital separation, system period, and orbital velocities.
  As in Tab.~3, bold values for the initial masses identify the
  component which ends its evolution first, typically by
  a supernova explosion.
  Both components of system No.~43 explode simultaneously, within the
  accuracy of the computation.}  
\begin{tabular}{l|l l l|r r r r r c c r r c c r c c }
\hline 
No. & M$_1$ & M$_2$ & P$_\mathrm{i}$ & M$^\prime_\mathrm{1}$ & M$^\prime_\mathrm{2}$ & 
q$^\prime$ & L$^\prime_\mathrm{1}$ & 
L$^\prime_\mathrm{2}$ & T$^\prime_\mathrm{eff,1}$ & T$^\prime_\mathrm{eff,2}$ & R$^\prime_1$ & R$^\prime_2$  & 
R$^\prime_2$ & a$^\prime$ & P$^\prime$ & v$^\prime_\mathrm{1}$ & v$^\prime_\mathrm{2}$  \\
 ~ & \multicolumn{2}{c}{$\Msun$} & d & \multicolumn{2}{c}{$\Msun$} & ~ & \multicolumn{2}{c}{log($\Lsun$)} &
 \multicolumn{2}{c}{$10^3 \mathrm{K}$} & \multicolumn{2}{c}{$\rso$} & R$_{\rm Roche}$ & $\rso$ & d & \multicolumn{2}{c}{km/s} \\
\hline
1     & 12 & {\bf 11.5} & 2.5 & 1.42 & 21.5 & 15.1 & 2.98 & 4.99 & 56   & 25   & 0.3 & 16.5 & 0.06 & 433 & 218 & 94   &  6.2 \\
2     & {\bf 12} & 11   & 3   & 2.44 & 20.4 &  8.4 & 3.93 & 4.74 & 63   & 33   & 0.8 & 7.2  & 0.07 & 170 & 54  & 143  & 17.1 \\
3     & {\bf 12} & 11   & 6   & 2.38 & 20.4 &  8.6 & 3.82 & 4.79 & 65   & 35   & 0.6 & 6.8  & 0.04 & 282 & 115 & 111  & 13.0 \\
10    & 12 & {\bf 10.5} & 2.5 & 1.42 & 20.7 & 14.6 & 2.97 & 4.87 & 56   & 34   & 0.3 & 7.9  & 0.03 & 390 & 190 & 97   &  6.7 \\
12    & {\bf 12} & 10   & 16  & 2.43 & 19.4 &  8.0 & 3.81 & 4.71 & 64   & 34   & 0.6 & 6.6  & 0.02 & 472 & 255 & 83   & 10.4 \\
15    & 12 & {\bf 9.5}  & 2.5 & 1.41 & 19.7 & 14.0 & 2.96 & 4.79 & 56   & 33   & 0.3 & 7.5  & 0.03 & 351 & 166 & 100  &  7.1 \\
16    & {\bf 12} & 9.5  & 3.5 & 2.45 & 18.9 &  7.7 & 3.94 & 4.63 & 62   & 33   & 0.8 & 6.3  & 0.07 & 158 & 50  & 142  & 18.4 \\
18    & {\bf 12} & 9    & 3.5 & 2.32 & 18.5 &  8.0 & 3.74 & 4.63 & 66   & 33   & 0.6 & 6.5  & 0.07 & 163 & 53  & 138  & 17.4 \\
19    & {\bf 12} & 9    & 6   & 2.36 & 18.5 &  7.8 & 3.81 & 4.63 & 65   & 33   & 0.6 & 6.2  & 0.04 & 227 & 87  & 117  & 15.0 \\
23    & 12 & {\bf 8.5}  & 2.5 & 1.43 & 18.8 & 13.1 & 2.97 & 4.72 & 57   & 33   & 0.3 & 6.8  & 0.04 & 301 & 135 & 105  &  8.0 \\
24    & 12 & {\bf 8}    & 2   & 1.12 & 18.5 & 16.5 & 2.64 & 4.84 & 48   & 32   & 0.3 & 8.5  & 0.04 & 373 & 189 & 94   &  5.7 \\
27    & {\bf 12} & 8    & 6   & 2.37 & 17.5 &  7.4 & 3.79 & 4.55 & 65   & 33   & 0.6 & 5.8  & 0.05 & 197 & 72  & 122  & 16.5 \\
31    & 12 & {\bf 7.5}  & 2.5 & 1.45 & 17.8 & 12.3 & 2.99 & 4.64 & 57   & 33   & 0.3 & 6.3  & 0.04 & 249 & 104 & 112  &  9.1 \\
 ~    &  ~ &  ~         &  ~  & ~    &  ~   &  ~   &  ~   &  ~   & ~    & ~    & ~   & ~    &  ~   &  ~  &  ~  &      &  ~   \\
34    & {\bf 16} & 15.7 & 3.2 & 2.79 & 27.8 & 10.0 & 3.88 & 5.14 & 76   & 37   & 0.5 & 9.3  & 0.07 & 234 & 75  & 143  & 14.4 \\
35    & {\bf 16} & 15.7 & 6   & 3.49 & 27.5 &  7.9 & 4.17 & 5.14 & 80   & 39   & 0.6 & 8.3  & 0.06 & 257 & 86  & 134  & 17.0 \\
37    & {\bf 16} & 15   & 8   & 3.63 & 26.7 &  7.4 & 4.20 & 5.10 & 80   & 36   & 0.7 & 8.9  & 0.06 & 275 & 96  & 128  & 17.3 \\
38    & {\bf 16} & 15   & 9   & 3.64 & 26.7 &  7.3 & 4.22 & 5.10 & 80   & 38   & 0.7 & 8.1  & 0.05 & 295 & 107 & 123  & 16.8 \\
42    & 16 & {\bf 14}   & 2.5 & 2.39 & 26.5 & 11.1 & 3.78 & 5.17 & 68   & 34   & 0.6 & 11.0 & 0.07 & 274 & 98  & 130  & 11.7 \\
43    & 16 & 14         & 3   & 2.52 & 26.6 & 10.6 & 3.75 & 5.10 & 72   & 36   & 0.5 & 9.0  & 0.06 & 280 & 101 & 128  & 12.2 \\
44    & {\bf 16} & 14   & 6   & 3.60 & 25.8 &  7.2 & 4.19 & 5.04 & 80   & 38   & 0.6 & 7.8  & 0.07 & 214 & 67  & 142  & 19.8 \\
47    & 16 & {\bf 13}   & 2.5 & 2.28 & 26.1 & 11.4 & 3.65 & 5.11 & 66   & 36   & 0.5 & 9.1  & 0.06 & 274 & 99  & 129  & 11.3 \\
49    & {\bf 16} & 13   & 3   & 2.64 & 25.3 &  9.6 & 3.84 & 5.07 & 74   & 35   & 0.5 & 9.1  & 0.08 & 204 & 64  & 146  & 15.2 \\
51    & {\bf 16} & 13   & 4   & 3.55 & 24.8 &  7.0 & 4.19 & 4.99 & 80   & 36   & 0.7 & 8.2  & 0.10 & 152 & 41  & 164  & 23.6 \\
55    & 16 & {\bf 12}   & 2   & 2.07 & 24.8 & 12.0 & 3.57 & 5.17 & 62   & 33   & 0.5 & 11.8 & 0.08 & 258 & 93  & 130  & 10.8 \\
56    & {\bf 16} & 12   & 4   & 3.55 & 23.9 &  6.7 & 4.17 & 4.93 & 80   & 37   & 0.6 & 7.0  & 0.09 & 141 & 37  & 168  & 24.9 \\
58    & 16 & {\bf 11}   & 2   & 2.00 & 24.1 & 12.1 & 3.56 & 5.10 & 62   & 33   & 0.5 & 10.8 & 0.07 & 249 & 89  & 131  & 10.8 \\
60    & {\bf 16} & 11   & 3   & 2.52 & 23.8 &  9.4 & 3.74 & 4.95 & 73   & 37   & 0.5 & 7.3  & 0.07 & 194 & 61  & 145  & 15.4 \\
61    & {\bf 16} & 11   & 3.2 & 2.81 & 23.3 &  8.3 & 3.87 & 4.93 & 77   & 36   & 0.5 & 7.4  & 0.09 & 151 & 42  & 162  & 19.5 \\
62    & {\bf 16} & 10   & 2.5 & 2.28 & 23.1 & 10.1 & 3.77 & 4.97 & 67   & 36   & 0.6 & 7.8  & 0.06 & 200 & 65  & 142  & 14.0 \\
  ~   &  ~ &  ~         &  ~  & ~    &  ~   &   ~  &  ~   &  ~   & ~    & ~    & ~   & ~    &    ~ &  ~  &  ~  &      &      \\
65    & 25 & {\bf 24}   & 3.5 & 5.15 & 40.6 &  7.9 & 4.53 & 5.54 & 98   & 38   & 0.6 & 13.7 & 0.14 & 179 & 41  & 196  & 24.8 \\
68    & 25 & {\bf 24}   & 5   & 7.28 & 39.2 &  5.4 & 4.85 & 5.45 & 106  & 39   & 0.8 & 11.9 & 0.15 & 149 & 31  & 205  & 38.1 \\
70    & 25 & {\bf 23}   & 4   & 5.31 & 39.4 &  7.4 & 4.56 & 5.49 & 99   & 39   & 0.6 & 12.4 & 0.13 & 172 & 39  & 196  & 26.4 \\
71    & 25 & {\bf 22}   & 2.5 & 4.80 & 39.6 &  8.3 & 4.45 & 5.55 & 91   & 37   & 0.7 & 14.2 & 0.15 & 171 & 39  & 198  & 24.0 \\
73    & {\bf 25} & 19   & 4   & 5.26 & 35.9 &  6.8 & 4.55 & 5.38 & 99   & 43   & 0.7 & 8.7  & 0.11 & 140 & 30  & 206  & 30.2 \\
74    & {\bf 25} & 16   & 4   & 5.22 & 33.3 &  6.4 & 4.54 & 5.29 & 99   & 41   & 0.6 & 8.8  & 0.14 & 118 & 24  & 215  & 33.8 \\
\hline
\end{tabular}
\end{table*}

While it is unlikely that any of the binary systems considered here
has an observable counterpart during the mass transfer --- with the
exception of the slow phase of the Case~A mass transfer --- they have
a relatively long-lived post-mass transfer stage. We emphasise that
this is true for those binary which survive the mass transfer, but
also for those which lead to mergers. Therefore, we characterise the
typical properties of systems after the first mass transfer phase in
Table~4. At this stage, the systems consist of a main sequence or
supergiant O or early~B star of 18...40$\mso$, and a helium star in
the mass range 1...7$\mso$. The orbital periods range from 24~d to
255~d.

An inspection of Table~4 indicates that it may be a challenging task to
try to detect systems like these observationally. A photometric
investigation will have limited success due to three factors.  First,
the luminosity ratio of both stars is large, mostly of the order
of~10, with the worst case of~160 (system No.~24) and the best of~4
(system No.~68). Second, the less luminous star has a much smaller
radius than the brighter star, which minimises the brightness
amplitude in case of an eclipse. Third, both stars underfill their
Roche volumes by a large factor, which means that they are likely to
radiate isotropically, and that eclipses are not likely.

We note that after core helium burning, all the helium stars below
$\sim 4\mso$ expand to giant dimensions and --- if the OB~star has not
yet turned into a supernova --- may even fill their Roche volumes (cf.
Sect.~3.2). As also their luminosities increase by up to a factor~10,
a photometric detection may indeed be more likely in this stage.
However, for the more massive helium stars (M$_{\rm 1}^\prime \simgr
3\mso$), this stage is very short lived ($\simle 0.1\,$Myr) compared
to their core helium burning life times ($\sim 1\,$Myr; cf. Table~2).
In systems with lower mass helium stars, the time scales are more
favourable: the 1.45$\mso$ helium star of system No.~31 burns helium
in its core for about 5~Myr, and its ensuing shell burning stage lasts
at least 1~Myr, possibly more depending on the mass loss rate during
this stage (cf. Sect.~3.1 and Table~1).

Table~4 also shows the orbital velocities of both components after the
mass transfer. While the helium stars have favourably large velocities
of the order of 100...200~km~s$^{-1}$, they might be unobservable
against the glare of their OB~companion. The latter moves with only
5...35~km~s$^{-1}$. While these velocities are small, it may not be
hopeless to try to detect these motions in OB~stars.

It is interesting to note that the assumption of conservative
evolution leads, in all respects, to system properties which are less
favourable for an observational detection of such systems, compared to
the assumption of mass- and angular momentum loss from the system.
This is demonstrated by the estimated outcome of our delayed contact
system No.~7 in Sect.~4.1.1, which results in a shorter orbital period
(7~d), correspondingly larger orbital velocities, and a mass- and
luminosity ratio closer to~1, compared to similar conservative
systems. This tendency can also be verified in the results obtained by
de~Loore \& De~Greve (1992) obtained with $\beta =0.5$, i.e. assuming
that half of the transferred mass leaves the system.

The best chances to find post-mass transfer systems like those of
Table~4 seems to be at the upper end of the considered mass range. Our
systems with initial primary masses of 25$\mso$ obtain the smallest
luminosity ratio L$_{\rm 2}^\prime$/L$_{\rm 1}^\prime$, the shortest
periods, and the largest orbital velocities. The helium star masses in
these systems are in the range 5...7$\mso$, which makes the stars
comparable to Wolf-Rayet stars, or, more specifically, WN~stars.
However, although we know 12~OB+WN systems with established mass
ratios and orbital periods (van der Hucht 2000), none of them fits to
our models. While 8 out of these 12 binaries have clearly much higher
initial system masses than those systems considered here, the
remaining four (WR3, WR31, WR97, and WR139; cf. van der Hucht 2000)
all have mass ratios of M$_{\rm OB}$/M$_{\rm WR} \simle 3$, and mostly
much shorter periods than our models.

There are two possible ways to interpret the disagreement between our
models and the observed OB+WN binaries. One possibility is that our
assumption of conservative evolution does not hold for systems with
primaries of initially $\sim 25\mso$. This can be motivated by the
increasing role of radiation pressure in higher mass systems (cf.,
Schuerman 1972; Vanbeveren 1978). On the other hand, strong
non-conservative effects in massive systems may result in an
uncomfortably large upper critical ZAMS mass for neutron star
formation of $\sim 40\mso$, as imposed by the massive X-ray binary
Wray~977 (Kaper et al. 1995, Ergma \& van den Heuvel 1998, Vanbeveren
1998b).

The other possibility involves to estimate the probability of finding
one system of the kind computed here among four OB+WR binaries. As
outlined in Sect.~5.1, the parameter range for conservative evolution
becomes narrower for higher masses. For our systems with 25$\mso$
primaries, contact is avoided for an initial period of 4~d, while 5~d
resulted already in contact (cf. systems No.~65...74 in
Table~3).Assuming initially an equal number of binaries per
log~P-interval, and an upper period limit for interaction of 2000~d,
results in a ratio of contact-free versus contact surviving binary
systems of 1/3.7 if all Systems with initial mass ratios q $\lesssim$
0.6 merge. If only the systems with initial mass ratios q $\lesssim$
0.6 and initial periods P $\lesssim$ 200~d merge this ratio becomes
1:6.6. Because it is not well known how much systems really merge ---
also an unknown fraction of the large period contact systems might
actually lead to a merger rather than to a OB+WR binary --- we assume
that the ratio is between 1:4 and 1:7. Thus the probability of finding
one suitable binary among four may be of the order of 1/2 or lower.
However, the survivors of the non-conservative contact evolution end
up as shorter period systems with a mass and luminosity ratio closer
to~one\footnote{This fits well to the properties of the WR binaries
  WR31, WR97, and WR139, while WR3 has most likely not experienced any
  mass transfer} than the contact-free systems, and are thus detected
much easier.

In summary, the failure to detect counterparts of our more massive
post-mass transfer systems is inconclusive. For our lower mass
systems, i.e. such with initial primary masses of 12$\mso$ and
16$\mso$, the chance to detect counterparts may even be lower than for
our higher mass systems (cf. Table~4). In fact, we do not know of {\em
  any} suitable OB~star with a 1...3.6$\mso$ helium star companion.
The object which comes closest to our models may be $\Phi\,$Persei, a
$\sim 9\mso$ B~star with a 53$\,$000$\,$K hot 1.1$\mso$ companion in a
127~d orbit (Gies et al. 1998). Its properties, and the fact that the
B~star is spinning rapidly, support actually the possibility of a
close-to-conservative evolution at least in this system.  A more
puzzling system which may be relevant in the context is
$\lambda\,$Sco, a $\sim 10.5\mso$ B~dwarf with a massive white dwarf
companion in a $\sim 6\,$d orbit (Bergh\"ofer et al. 2000).  Waters et
al. (1989) speculate that unseen companions in a number of Be stars
may actually be helium stars, in accordance with Pols et al. (1991)
and Portegies Zwart (1995).

The lack of OB+He star systems which are more massive than $\Phi\,$Per
but less massive than OB+WR systems is, anyhow, very remarkable. I.e.,
if the evolution of binaries with 12...16$\mso$ primaries would, for a
big fraction of the possible period range, proceed non-conservatively
and without leading to mergers, we would expect short period OB+He
star systems which are easier detectable, with a smaller mass- and
luminosity ratio compared to our models. Therefore, a non-detection of
such systems may be easiest understood if the conservative evolution
makes up for a large fraction of the non-merging interacting binaries
in the considered mass range.

Whichever way it is, there are two kinds of observations hinting for a
large number of OB+He star binaries waiting to be discovered. One is
the existence of massive X-ray binaries, which are likely evolving
from an OB+He star stage (e.g., Moers \& van den Heuvel 1989).  The
second is the relatively large number of Type~Ib/Ic supernovae, which
are mostly attributed to exploding helium stars with masses above
2$\mso$ (Podsiadlowski et al.  1992).  I.e., it might be a rewarding
project to attempt an observational search for helium star components
to OB stars. In that respect, Table~4 can give indications of the
expected parameters, with the remark that if mass transfer is partly
non-conservative, then mass ratio, luminosity ratio, period and
orbital velocities become more favourable for a detection.

Assuming a ratio of interacting binaries to single stars of 0.3, and
ignoring the unknown fraction of mergers among the binaries, implies
that 1 out of 3 OB stars in the considered mass range would, during
its life time, obtain a He~star companion. Depending of the
rejuvenation of the secondary, this phase may last a considerable
fraction, say one third, of the secondary's life time. This optimistic
estimate indicates that only one out of ten OB stars in the considered
mass range may have a helium star companion. However, these OB~stars
have a surface composition which is typically enriched in helium by
10...20\%, and which has a N/C-ratio of about one order of magnitude
above solar (cf. Sect.~3, and Wellstein \& Langer 2001). I.e., even
though also single stars may become helium- and nitrogen-rich during
their main sequence evolution due to rotational mixing (Heger \&
Langer 2000), a pre-selection of chemically enriched OB~stars may lead
to a large detection probability of helium star companions.

Another way to pre-select OB stars with potential helium star
companions is to identify OB~type blue stragglers in young clusters or
associations. The accretion of substantial amounts of matter by the
primary may lead it to achieve a mass which is already above the
turn-off mass of the cluster or associations. E.g., the 7.5$\mso$
Case~A secondary shown in Fig.~1 grows to about 18$\mso$ at an age of
13~Myr, which is the turn-off age for 13...14$\mso$. The 15$\mso$
Case~B primary, upon turn-off of its 16$\mso$ companion, even grows to
27$\mso$. The remaining main sequence life time of the secondaries
after accretion is of the order of 1~Myr, which may be long enough to
catch them during this stage. Furthermore, the accretion of angular
momentum may spin up the star, and lead to internal mixing processes 
and thereby to an overluminosity of the star with respect to its mass. 

In summary, enriched surface abundances, blue straggler nature,
overluminosity for the actual mass, and rapid rotation in an O~or
early B~star may be hints to a helium star companion.

\section{Summary and conclusions}

In this paper, we have, on the basis of evolutionary models with
up-to-date input physics, investigated the regime of contact-free
evolution in close binaries with initial primary masses in the range
12...25$\mso$, under the assumption that contact-free systems evolve
conservatively, i.e., that no matter except stellar winds leaves these
systems. Our results confirm earlier obtained ones qualitatively
(e.g., Pols 1994), but differ quantitatively and extend over a wider
parameter range.

We demonstrate that the number ratio of contact free Case~A-to-Case~B
systems is a strong function of the system mass. At the high mass end
(25$\mso$ initial primary mass), the contact-free Case~B channel is
closed completely, while contact-free Case~A systems exist for a wide
parameter range (cf. Table~3), although we find this range to depend
strongly on the semiconvective mixing efficiency (cf. Sect.~5, and
Wellstein \& Langer 1999). In contrast, the contact-free Case~A
channel becomes narrow at 12$\mso$ initial primary mass, whatever the
assumptions on semiconvective mixing are, while the contact-free
Case~B range becomes large (Fig.~12). At an initial primary mass of
16$\mso$, the situation is intermediate (Fig.~13).

We investigate the vastly different types of contact situations of
systems in the considered mass range, and determine the critical
periods and mass ratios to avoid contact (Sect.~5.1). We also estimate
the consequences of the various kinds of contact. As a new phenomenon,
we find that in what we call {\em delayed contact}, Case~B systems can
transfer the majority of the primary's envelope conservatively before
--- towards the end of the Case~B mass transfer --- contact and
possibly a common envelope evolution occur. We find the delayed
contact regime may lead, with increasing initial period, to a smooth
transition from conservative contact-free evolution to the classical
common envelope evolution, with the parameter~$\beta$, which describes
which fraction of the primary's envelope is accreted by the secondary,
varying smoothly from $\beta =0$ to $\beta =1$. We are able to compute
values of $\beta$ for our delayed contact systems, which we find in
the range $\beta$ = 0.4...0.7.

How far or close our model sequences actually are to the evolution of
binary systems in nature depends to a large extent on the question
whether contact-free systems do indeed evolve conservatively.  I.e.,
the accreting star is likely to be rapidly spun-up to close to
critical rotation (e.g., Packet 1981), and it depends on unknown
viscous time scales whether or not it will be able to continue to grow
in mass (Paczy\'nski 1991). This question can in principle be answered
by comparing our post-mass transfer models with corresponding observed
binaries. However, as outlined in Sect.~6, even though virtually no
observed counterparts to our post-mass transfer models --- i.e., no
OB+He star systems with large mass ratios and orbital periods --- are
known, we could not conclude that conservative evolution in massive
binaries is not realised in nature, for two reasons. First,
identifying rather far and faint helium star companions to bright OB
stars may be too difficult. Second, non-conservative evolution
predicts even easier detectable OB+He star systems but, again, no such
systems are known, at least not corresponding to primary masses
between 12 and 20$\mso$ --- while three known OB+WN binaries might
indeed be the result of non-conservative binary evolution with
primaries above~20$\mso$.

We want to emphasise that an answer to the question whether massive
binaries can evolve conservatively is important to understand the late
evolutionary stages: e.g., Type~Ib/Ic supernovae, massive X-ray and
black hole binaries, and $\gamma$-ray bursts due to massive binaries.
In principle one may constrain the conservativity of massive binaries
by observations of the advanced stages. However, assumptions on even
more uncertain processes have to be made in such attempts --- e.g., on
common envelope evolution, neutron star birth kicks, etc. --- which
strongly limits the success of such approaches. Furthermore, often,
e.g. in population synthesis studies, only statistical information is
obtained. For example, Dalton \& Sarazin (1995) find the period
distribution of massive X-ray binaries consistent with an average loss
of $\sim 70$\% of the transferred envelope mass. However, this can
neither support nor defeat the existence of conservative evolution for
contact-free massive binaries.

The very best way to constrain the first major mass transfer phase in
massive close binaries would indeed be to identify a sample of helium
star companions to OB stars. We repeat that those are predicted by
conservative {\em and} by non-conservative binary evolution models,
but with different periods and mass ratios.  They are also indirectly
proven to exist abundantly by the relatively large number of
Type~Ib/Ic supernovae and massive X-ray binaries.  In Sect.~6, we
identify several ways to pre-select OB star candidates with an
improved probability for a helium star companion.

\begin{acknowledgements}
  We are grateful to Doug Gies,
  Lex Kaper, Jerry Orosz, and Philipp Podsiadlowski for many
  helpful discussions and suggestions and to Karel van der Hucht for
  providing us with results prior to publication.  This work has been
  supported by the Deutsche Forschungsgemeinschaft through grants
  La~587/15 and La~587/16.
\end{acknowledgements}

\end{document}